\documentclass[fleqn,usenatbib]{mnras}

\usepackage{newtxtext}

\usepackage[dvipsnames]{xcolor}

\usepackage[T1]{fontenc}
\usepackage{ae,aecompl}

\usepackage{graphicx}	
\usepackage{amsmath}	
\usepackage{amssymb}	

\def\gtorder{\mathrel{\raise.3ex\hbox{$>$}\mkern-14mu
     \lower0.6ex\hbox{$\sim$}}}
\def\ltorder{\mathrel{\raise.3ex\hbox{$<$}\mkern-14mu
     \lower0.6ex\hbox{$\sim$}}}

\title[Non-thermal Sgr A* X-ray flares from a magnetically arrested disc]{Sgr A* X-ray flares from non-thermal particle acceleration in a magnetically arrested disc}

\author[N. Scepi et al.]{
Nicolas Scepi,$^{1}$\thanks{E-mail: nisc6580@colorado.edu}
Jason Dexter$^{1,2}$
and Mitchell C. Begelman,$^{1,2}$
\\
$^{1}$JILA, University of Colorado and National Institute of Standards and Technology, 440 UCB, Boulder, CO 80309-0440, USA \\
$^{2}$Department of Astrophysial and Planetary Sciences, University of Colorado, 391 UCB, Boulder, CO 80309-0391, USA}

\date{Accepted XXX. Received YYY; in original form ZZZ}

\pubyear{2020}

\begin{document}
\label{firstpage}
\pagerange{\pageref{firstpage}--\pageref{lastpage}}
\maketitle

\begin{abstract}
Sgr A* exhibits flares in the near-infrared and X-ray bands, with the luminosity in these bands increasing by factors of $10-100$ for $\approx 60$ minutes. One of the models proposed to explain these flares is  synchrotron emission of non-thermal particles accelerated by magnetic reconnection events in the accretion flow. We use the results from PIC simulations of magnetic reconnection to post-process 3D two-temperature GRMHD simulations of a magnetically arrested disc (MAD).  We identify current sheets, retrieve their properties, estimate their potential to accelerate non-thermal particles and compute the expected non-thermal synchrotron emission. We find that the flux eruptions of MADs can provide suitable conditions for accelerating non-thermal particles to energies $\gamma_e \lesssim 10^6$ and producing simultaneous X-ray and near-infrared flares. For a suitable choice of current-sheet parameters and a simpified synchrotron cooling prescription, the model can simultaneously reproduce the quiescent and flaring X-ray luminosities as well as the X-ray spectral shape. While the near-infrared flares are mainly due to an increase in the temperature near the black hole during the MAD flux eruptions, the X-ray emission comes from narrow current sheets bordering highly magnetized, low-density regions near the black hole and equatorial current sheets where the flux on the black hole reconnects. As a result, not all infrared flares are accompanied by X-ray ones. The non-thermal flaring emission can extend to very hard  ($\lesssim 100$ keV) X-ray energies.
\end{abstract}

\begin{keywords}
Galaxy: centre -- accretion discs -- variability -- radiation mechanisms: non-thermal -- acceleration of particles -- magnetic field 
\end{keywords}


\makeatletter
\let\origsection\section
\renewcommand\section{\@ifstar{\starsection}{\nostarsection}}

\newcommand\nostarsection[1]
{\sectionprelude\origsection{#1}\sectionpostlude}

\newcommand\starsection[1]
{\sectionprelude\origsection*{#1}\sectionpostlude}

\newcommand\sectionprelude{%
  \vspace{-1mm}
}

\newcommand\sectionpostlude{%
  \vspace{-1mm}
}
\makeatother
\section{Introduction}
First discovered in the radio \citep{balick1974}, the supermassive black hole at the center of our Galaxy, Sagittarius A* (Sgr A*), is now routinely observed in the near-infrared (NIR) \citep{genzel2003}, X-ray \citep{baganoff2001} and gamma-ray bands \citep{mayer1998}. It is now accepted that the emission from Sgr A* comes from an optically thin, hot accretion flow around a black hole of mass $4\times10^6 M_\odot$ \citep{ghez2005,gillessen2017,gravity2018a}. With a bolometric luminosity of $\approx 5\times 10^{35}\:\mathrm{erg\:s^{-1}}$ \citep{bower2019}, Sgr A* is accreting at a very sub-Eddington rate and is the best laboratory to study radiatively inefficient accretion flows \citep{yuan2014}. 

In its quiescent state, the spectrum of Sgr A* typically extends from radio to NIR wavelengths and is thought to originate from synchrotron emission of a population of relativistic, thermal electrons in a magnetic field of $\approx 100$ G \citep{bower2019}. In quiescence, the level of X-rays is lower than any other low-luminosity AGN and is believed to originate from thermal bremsstrahlung emission near the Bondi radius \citep{quataert2002,baganoff2003}.

Several times a day, Sgr A* shows events of strong variability in the NIR and X-ray bands, called flares \citep{genzel2003,ghez2005,baganoff2001,neilsen2013}. During a flare the luminosity in the X-rays and in the NIR can rise by a factor as large as $\approx100$ \citep{do2019,haggard2019}. Since the sub-millimeter emission does not show simultaneous, large-amplitude variability \citep[e.g.,][]{eckart2006b,yusefzadeh2008,marrone2008}, it is usually assumed that these events are due to a separate, possibly non-thermal, population of electrons \citep{markoff2001}.  Given that the NIR emission is highly polarized \citep{eckart2006a}, it is further assumed that these electrons radiate their energy via synchrotron emission \citep{dodds2009,ponti2017}, although a synchrotron self-Compton origin for the X-rays may also be possible  \citep[e.g.,][]{eckart2012,dibi2016}.

Recently, the GRAVITY instrument was able to resolve the motion of the NIR centroid during a flaring event, showing a clockwise, continuous rotation consistent with a region of emission located at a few gravitational radii, $r_g$, from the central black hole \citep{gravity2018b}. This result suggests that the particles responsible for the flaring are accelerated in a compact, rotating region near the black hole.

One scenario to produce these features is the presence of magnetic reconnection events in a collisionless, hot plasma rotating around the black hole. The magnetic reconnection scenario is supported by the results of particle-in-cell (PIC) simulations showing that the parallel electric field generated during reconnection events can accelerate particles to very high energies, producing a power-law distribution of electrons with hard indices \citep{bessho2012,cerutti2012,cerutti2013,kagan2013,guo2014,sironi2014,guo2016,werner2016,werner2018,werner2019}. Synchrotron emission from a non-thermal distribution of electrons of index $\approx2$ could explain the flaring spectrum of Sgr A* \citep{dodds2009}. 

Although PIC simulations are able to resolve the plasmoid dynamics and the subsequent particle acceleration, they are restricted to very limited computational domains.  Current sheets are apparent at all times in general relativistic magnetohydrodynamic (GRMHD) simulations  \citep[e.g.,][]{gammie2003,hirose2004}. When using explicit resistivity, they reconnect into plasmoids \citep{ripperda2020}, and may provide potential locations for magnetic reconnection flares. However, the details of non-thermal particle acceleration require a kinetic approach, such as that provided by PIC.

Nevertheless, much effort has been put into modeling Sgr A* from first-principles GRMHD simulations. For the low Eddington ratio of Sgr A*, radiative cooling can be safely neglected \citep{dibi2012,ryan2017} and observables can be calculated in post-processing. Since the plasma is collisionless, electrons and ions are not necessarily thermally coupled. Radiative transfer calculations must either assign an ion to electron temperature ratio \citep[e.g.,][]{moscibrodzka2009,dexter2010,shcherbakov2012,chan2015} or evolve the electron entropy separately \citep{ressler2015,ryan2017,chael2018,dexter2020}. In the latter case, the total heating rate is divided into electrons and ions using some sub-grid prescription based on kinetics calculations  \citep{howes2010,rowan2017,werner2018,kawazura2019}. In almost all cases, the electron distribution function has been assumed to be purely thermal.

Radiative models from GRMHD simulations can generically reproduce the shape of the Sgr A* spectral peak in the sub-millimeter and its compact source size \citep{doeleman2008}. Turbulence driven by the magnetorotational instability \citep[MRI,][]{mri,balbus98} can also explain the $\simeq 30\%$ submm flux density fluctuations \citep{dexter2009}. Current GRMHD models have more difficulty in reproducing the large-amplitude NIR and X-ray flares. In the NIR, promising scenarios include misaligned accretion discs \citep{dexter2013,white2021}, gravitational lensing events \citep{chan2015flare}, electron heating at the jet wall near the event horizon \citep[e.g.,][]{ressler2017}, and magnetic eruptions in strongly magnetized discs \citep{dexter2020b,porth2021}. Efforts to introduce non-thermal  electrons to model flares \citep[e.g.,][]{ball2016,chatterjee2020,petersen2020} show some promise. However, so far they have difficulty  reproducing the large contrast in X-ray luminosity between flares and quiescence, and the lack of corresponding increases in the sub-millimeter luminosity.

Magnetically arrested discs \citep[MAD,][]{bisnovatyi1974} are characterized by stochastic eruption events of magnetic flux from near the black hole \citep[e.g.,][]{igumenshchev2008,tchekhovskoy2011}. \citet{dexter2020b} found that GRMHD models of MADs with self-consistent electron heating provide a promising scenario for explaining Sgr A* flares. The infrared flare luminosity, timescale, and recurrence time from magnetic flux eruptions near the black hole are broadly consistent with those observed. Electron heating in rotating interfaces between magnetically dominated bubbles and high density gas caused rotating flaring emission regions, whose photocenter and linear polarization evolution seem promisingly similar to those found in GRAVITY observations \citep{gravity2018b,gravity2020,gravity2020pol}. In these simulations, energy disspation occurs from (numerical) magnetic reconnection, and the current sheets involved seem promising for accelerating particles to high energy \citep{porth2021}. However, that work assumed a purely thermal electron distribution.

Here, we assess the potential of magnetic eruption events in MADs for producing luminous X-ray flares due to synchrotron radiation from a non-thermal distribution of electrons accelerated by magnetic reconnection.  In \S\ref{sec:methods}, we  introduce our methods for identifying reconnecting current sheets in ideal GRMHD, retrieving their properties, and estimating their potential for accelerating non-thermal particles. We calculate approximate spectra and light curves from the resulting non-thermal distribution function in \S\ref{sec:results}, and show that luminous X-ray flares are likely to result from eruption events. For some parameter choices, the resulting quiescent and flaring luminosities match those observed from Sgr A*. In \S\ref{sec:discussion}, we discuss the implications of our results and directions for future work.

\section{Methods}\label{sec:methods}
In this section we describe our method to identify current sheets, to retrieve the properties of these current sheets, to assign a non-thermal particle distribution at each cell and to estimate the subsequent synchrotron emission coming from each cell.

\subsection{Numerical set-up}

We use the results of a 3D MAD GRMHD simulation and a standard accretion and normal evolution (SANE) GRMHD simulation, both already presented in \cite{dexter2020}. We will present the general properties of the simulations but for more information the reader can refer to \cite{dexter2020}. The GRMHD simulations were performed with the public code \textsc{harmpi}\footnote{https://github.com/atchekho/harmpi} \citep{tchekhovskoy2019} that includes a scheme to evolve electron internal energy densities along with that of the MHD fluid \citep{ressler2015}. We use an adiabatic equation of state with a relativistic adibatic index of $4/3$ for the electrons and a non-relativistic adiabatic index of $5/3$ for the MHD fluid. The prescription for electron heating that is used for the MAD and SANE simulation derives from the PIC reconnection study of \cite{werner2018}.

The simulations were initialized from a Fishbone-Moncrief torus \citep{fishbone1976} with an inner radius of 12 $r_g$, pressure maximum at 25 $r_g$, and a black hole spin parameter of $a=0.9375$. The simulation grid uses coordinates based on a spherical-polar Kerr-Schild metric, which are distorted in the $\theta$-direction in order to better resolve the equatorial inner accretion flow and extended relativistic jet. The radial coordinate is logarithmically spaced. The grid resolution is $320\times256\times160$ in the $r,\theta$ and $\phi$ directions, respectively. The magnetic field configuration is initialized as a single poloidal field loop, whose amplitude is chosen so that $\mathrm{max}(p_g)/\mathrm{max}(p_B)=100$, where $p_g$ is the gas pressure and $p_B\equiv b^\mu b_\mu /2$ the magnetic pressure and whose radial profile is chosen so as to produce either a MAD or a SANE.

The MAD simulation has been run for $9\times10^4\:r_g/c$ and has established an inflow equilibirum up to $\approx90\:r_g$ by the end of the simulation. The simulation becomes MAD around $6\times10^3\:r_g/c$. The SANE simulation has been run for $19\times10^4\:r_g/c$ so that it also achieves an inflow equilibrium up to $\approx90\:r_g$.

\subsection{Identification of current sheets}\label{sec:id_current_sheets}
In ideal MHD, the electromagnetic field tensor can be expressed as 
\begin{equation}
F^{\mu\nu} = \epsilon^{\mu\nu\kappa\lambda}u_\kappa b_\lambda
\end{equation}
where $u_\mu$ is the covariant four-velocity, $b_\mu$ is the covariant 4-magnetic field, $\epsilon^{\mu\nu\kappa\lambda}=(-1/\sqrt{-g})[\mu\nu\kappa\lambda]$, with $[\mu\nu\kappa\lambda]$  the completely antisymmetric symbol and $g=\mathrm{det}(g_{\mu\nu})$ \citep{gammie2003}. The 4-current is expressed as

\begin{equation}
    \mathcal{J}^\mu={F^{\mu\nu}}_{;\nu}
\end{equation}
where ${}_{;\mu}$ represents the covariant derivative. Using the symmetry of the Christoffel symbol about its lower indices and the anti-symmetry of the electromagnetic field tensor gives
\begin{equation}\label{eq:current1}
    \mathcal{J}^\mu=   \frac{1}{\sqrt{-g}}\partial_\nu(\sqrt{-g} F^{\mu \nu}).
\end{equation}
To obtain a reasonable estimate of the current without taking approximate time derivatives from our data, we transform the electromagnetic tensor to the rest frame of the fluid where the electric current vanishes. We then approximate the derivatives in the rest frame with derivatives in the coordinate frame to approximate the current in the rest frame as
\begin{equation}\label{eq:current}
    \mathcal{\bar{J}}^i \approx   \frac{1}{\sqrt{-g}}\partial_j(\sqrt{-g} \bar{F}^{ij}),
\end{equation}
where barred quantities are computed in the rest frame. We verified on one snapshot that computing the current from Eq. \ref{eq:current1} on the fly in our simulation only leads to a $20\%$ difference in the norm of the current compared to the approximate value from Eq. \ref{eq:current}. We then introduce the parameter
\begin{equation}
    \mathcal{C}\equiv \frac{|\mathcal{\bar{J}}|\delta}{|b|},
\end{equation}
where $|\mathcal{\bar{J}}|=\sqrt{\mathcal{\bar{J}}^\mu \mathcal{\bar{J}}}$, $|b|=\sqrt{b^\mu b_\mu}$ and $\delta$ is the size of a cell in the rest frame \citep{bodo2021}. Since high values of $\mathcal{C}$ correspond to zones of high current and low magnetic field, we use this parameter to identify potentially reconnecting current sheets in our simulations. \cite{bodo2021} verified that the current sheets found with this method are very similar to those found with the more advanced algorithm developed in \cite{zhdankin2013}. 

\begin{figure}
\includegraphics[width=80mm]{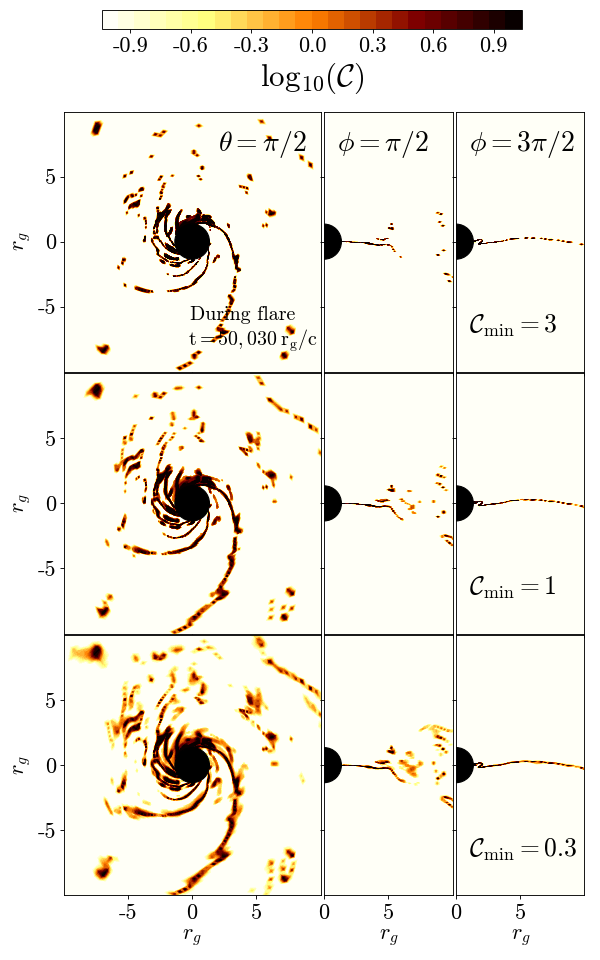}
\caption{Selection of current sheets during a flaring event at t=50,030 $r_g/c$.  Current sheets are found both during and outside flares. From left to right: equatorial cut at $\theta=\pi/2$ and poloidal cuts at $\phi=\pi/2$ and $\phi=3\pi/2$. From top to bottom, we used $\mathcal{C}_{\rm min}=3,1$ and 0.3.}
\label{fig:C_min_flare}
\end{figure}

\begin{figure}
\includegraphics[width=80mm]{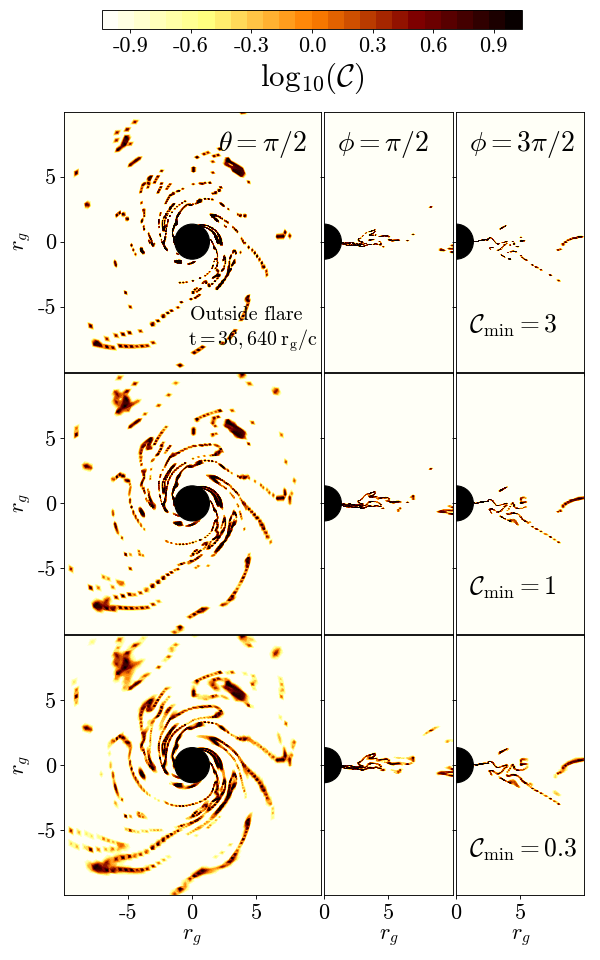}
\caption{Selection of current sheets during a  non-flaring epoch at t=36,640 $r_g/c$.  We see similar patterns of current sheets in both the flaring and non-flaring cases. From left to right: equatorial cut at $\theta=\pi/2$ and poloidal cuts at $\phi=\pi/2$ and $\phi=3\pi/2$. From top to bottom, we used $\mathcal{C}_{\rm min}=3,1$ and 0.3}
\label{fig:C_min_noflare}
\end{figure}

Figures \ref{fig:C_min_flare} and \ref{fig:C_min_noflare} show equatorial (left) and poloidal (two right) cuts of $\mathcal{C}$ where we used different thresholds, $\mathcal{C}_\mathrm{min}=0.3,1$ or 3, during and outside a magnetic eruption event. In both cases, we identify thin elongated structures that resemble current sheets. The lower the threshold $\mathcal{C}$, the thicker and the more numerous become the current sheets. Our method identifies similar patterns of current sheets during and outside eruption events. However, as we will see in \S\ref{sec:current_sheet_emission}, the current sheet properties are very different in these two cases.

\subsection{Properties of the current sheets and non-thermal particle distribution}\label{sec:NTPA}
To assess the potential of the currents sheets as non-thermal particle accelerators, we need to systematically retrieve their physical properties. Important quantities in reconnection studies are the upstream cold magnetization, $\sigma_\mathrm{up}=b^2/\rho$, and the upstream plasma $\beta$-parameter, $\beta_\mathrm{up}$, as well as the local number density of electrons, $n_e$, the local electron temperature, $T_e$, and the local magnetic field strength, $B$ \citep{werner2018}. We define the local properties as the properties of a given cell in our GRMHD simulation and the upstream properties as the properties of the surrounding cells. To retrieve the upstream properties we draw a sphere of 5 cells \footnote{Our choice of 5 cells is the minimum number of cells required to produce enough hard non-thermal power-law emission. For less than 5 cells the emission becomes steeper due to the low magnetic field inside the current sheets. For more than 5 cells the non-thermal emission does not change much with an X-ray flux that is only 2 times higher for 20 cells than for 5 cells at $t=50,300\:r_g/c$ and for $\mathcal{C}_\mathrm{min}=1$.} around each cell and look for the most magnetized cell, which we then assume to give the magnetization of the upstream region.  

We then use the upstream properties and local properties of the current sheets to assign to each cell a modified electron $\kappa$-distribution. The $\kappa$-distribution is a hybrid distribution composed of a thermal bulk of electrons at low energy and a non-thermal power-law tail at high energy, which is, in the relativistic regime, expressed as 
\begin{equation}
\frac{d n_e}{d\gamma} = N \gamma(\gamma^2-1)^{1/2}\left(1+\frac{\gamma-1}{\kappa w}\right)^{-(\kappa+1)}\times e^{-\gamma/\gamma_\mathrm{cut}} .
\end{equation}
Our modification consists of an exponential cutoff at high energies, $\gamma_\mathrm{cut}$, the value of which depends on cooling as explained in \S\ref{sec:cooling}. Here $n_e$ is the number density of electrons, $\gamma$ is the Lorentz factor, $N$ is a normalization factor, $\kappa=p+1$, $p$ is the index of the power-law tail, $\gamma^{-p}$, and $w$ is a parameter defining the width of the distribution, which tends to $\theta_e$, the dimensionless electron temperature $k_BT_e/m_e c^2$, as the power-law component becomes very steep. 

Using the results of PIC simulations, we can constrain the value of $p$ from the upstream and local properties of the current sheets. We use the following scaling law from \cite{ball2018} and \cite{werner2018},
\begin{equation}
    p = A_p + B_p \mathrm{tanh}(C_p\beta_\mathrm{up})
\end{equation}
 where $A_p=1.8+0.7/\sqrt{\sigma_\mathrm{up}}$, $B_p= 3.7\sigma_\mathrm{up}^{-0.19}$, $C_p=23.4\sigma_\mathrm{up}^{0.26}$.
 
 Once we constrain the power-law index of the $\kappa$-distribution, we need to constrain the two remaining parameters which are the normalization factor $N$ and the width of the distribution $w$. To do that, we use the two first moments of the electron energy distribution:
 \begin{gather}
     n_e = \int_{1}^{\infty}  \frac{d n_e}{d\gamma} d\gamma, \\
     u_e = \int_{1}^{\infty} \gamma\frac{d n_e}{d\gamma} d\gamma,
 \end{gather}
 
 \noindent where $n_e = n_i = \rho/m_p$ from the fluid mass density and $u_e$ is the self-consistently evolved electron internal energy density.

\subsection{Synchrotron emission}
We estimate the synchrotron emission of a thermal and a $\kappa$-distribution of electrons by using the fitting formul\ae\ of \cite{pandya2016}. The synchrotron emissivity in vacuum can be written as 
\begin{equation}
    j_S = \frac{n_e e^2 \nu_c}{c}J_S\left(\frac{\nu}{\nu_c},\theta\right)
\end{equation}
where the formula for $J_S$ depends on the electron distribution, $\nu_c\equiv e |b|/(2\pi m_e c)$ is the electron cyclotron frequency and $\theta$ is the angle between the observer and the magnetic field. 

For a relativistic thermal distribution, \citet{pandya2016} find 
\begin{equation}
    J_\mathrm{S} = \frac{\sqrt{2}\pi}{27}e^{-{X^{1/3}}}\sin(\theta)(X^{1/2}+2^{11/12}X^{1/6})^2
\end{equation}
where $X\equiv \nu/\nu_s$ and $\nu_s\equiv (2/9)\nu_c\theta_e^2\sin(\theta)$. 

For a $\kappa$-distribution, they find
\begin{equation}\label{eq:kappa_emiss}
    J_\mathrm{S} =  (J_\mathrm{S,low}^{-x}+J_\mathrm{S,high}^{-x})^{-1/x}
\end{equation}
where 
\begin{align}
    J_\mathrm{S,low} =&  X_{\kappa}^{1/3}\sin(\theta)\frac{4\pi\Gamma(\kappa-4/3)}{3^{7/3}\Gamma(\kappa-2)}\\
    J_\mathrm{S,high} =& X_{\kappa}^{-(\kappa-2)/2}\sin(\theta)3^{(\kappa-1)/2}\frac{(\kappa-2)(\kappa-1)}{4}\times\\
    &\Gamma\left(\frac{\kappa}{4}-\frac{1}{3}\right)\Gamma\left(\frac{\kappa}{4}+\frac{4}{3}\right)\nonumber,
\end{align}
 $X_\kappa\equiv \nu/\nu_\kappa$, $\nu_\kappa\equiv \nu_c(w \kappa)^2$, $x = 3\kappa^{-3/2}$, and $\Gamma$ is the gamma function.\\

Once we have computed the emissivities, we compute the thermal absorptivity, $\alpha_S\equiv j_S/B_\nu$, where $B_\nu\equiv(2h\nu^3/c^2)[\exp{(h\nu/kT_e)}-1]^{-1}$ is the Planck function.\footnote{We use the Kirchoff's law to compute the absorptivity since non-thermal electron emission is only important at high frequencies where the plasma is optically thin.} We then integrate the radiative transfer equation along the $\theta$-coordinate \footnote{Since most of the emission comes from the bulk of the disc, integrating along $\theta$ is almost equivalent to integrating along straight vertical lines.} to compute the luminosity, 
\begin{multline}
   L_\nu = \int_{\mathrm{ergo}}^{50 r_g}\int_0^{\tau_\nu}\int_0^{2\pi} 
   e^{-(\tau_\nu(r,\pi,\phi)-\tau_\nu(r,\theta,\phi))} \times\\
   j_S\times g_\mathrm{redshift}^3 \sqrt{-g}dr d\theta d\phi.
\end{multline}
where $\tau_\nu\equiv \int_{0}^{\theta}\alpha_S ds$ is the local total optical depth in the disc along $\theta$, $ds=\sqrt{g_{\theta\theta}}d\theta$, $g_\mathrm{redshift}\equiv\sqrt{A\Sigma/\Delta}$ is the gravitational redshift between the emitted frequency in the rest-frame and the frequency received by a distant static observer \citep{viergutz1993} with $A=(r^2+a^2)^2-a^2\Delta \sin^2\theta$, $\Sigma=r^2+a^2 \cos^2\theta$ and $\Delta=r^2-2rM+a^2$ and $M$, the mass of the central object, is set to 1 in this formula. In order to further take into account the gravitational effects, we also remove all the emission coming from the ergosphere since the redshift effects are usually large in this region, making it very dim. Note that, as in \citep{dexter2020}, we ignore the emission coming from the regions where $\sigma>1$ since these regions are likely to be affected by numerical floors.

This simple integration allows us to take into account the synchrotron self-absorption (this is only important at low frequencies since the plasma is optically thin at frequencies higher than $\approx 10^{12}$ Hz) as well as part of the general relativistic effects. The approximation is most suitable for low inclination angle (almost face-on), since we neglect Doppler beaming and gravitational lensing effects. We show below that the induced errors in the flared thermal NIR emission, although it originates from very close to the black hole where gravitational effects are important, are of order unity, while here we are mainly interested in order-of-magnitude estimates of the flaring and quiescent non-thermal X-ray luminosity.

\subsection{Effect of cooling on the particle distribution}\label{sec:cooling}
Cooling can affect the particle distribution in two ways. Very strong cooling can prevent the acceleration of high energy particles in the reconnecting current sheet in the first place. This happens if the synchrotron cooling timescale is shorter than the acceleration timescale due to the non-ideal electric field in the current sheet. The timescales are equal for an electron Lorentz factor $\gamma_\mathrm{rad}\equiv \sqrt{3 e E_\mathrm{rec}/(4 \sigma_T U_B)}$ where $E_\mathrm{rec}\approx0.1 |b|$ is the non-ideal electric field \citep{uzdensky2011}, $e$ is the electron charge, $\sigma_T$ is the Thomson cross-section and $U_B$ is the magnetic energy density. Based on the strong inverse-Compton cooling PIC simulations by \cite{werner2019}, we use a precription for the cutoff frequency at each cell that depends on the value of $\gamma_\mathrm{rad}$,
\begin{equation}
    \gamma_\mathrm{cut} = \frac{\gamma_\mathrm{rad}}{\sqrt{1+0.0625(\gamma_\mathrm{rad}/(m_p/m_e) \sigma_\mathrm{up})^2}}.
\end{equation}
When cooling is weak, $\gamma_\mathrm{cut} = 4 (m_p/m_e)\sigma_\mathrm{up}$ \citep{werner2016}. When cooling is strong, $\gamma_\mathrm{cut}=\gamma_\mathrm{rad}$. We find that we never actually reach the strong cooling regime and that the cutoff is located at $ \approx 4 (m_p/m_e)\sigma_\mathrm{up}$. Note that our choice of $\gamma_\mathrm{cut}$ for weak cooling is rather conservative since PIC simulations show that particles can be accelerated to higher energies in plasmoids at later times, with the cutoff frequency going as the square root of time \citep{petropoulou2018,hakobyan2021}.

The second way that cooling can affect the particle distribution is by cooling the particles after they have been accelerated. In particular, it is well known that if particles do not have time to cool before they get advected onto the black hole or leave the system (weak cooling), then the particle distribution is similar to the spectrum of injected particles. However, if particles can cool before they leave the system (strong cooling), the power-law index of the particle distribution is steepened by 1 compared to the weak cooling case \citep{blumenthal1970}. To assess whether the cooling can significantly affect the particle distribution or not we compare the synchrotron cooling time, $t_\mathrm{synch}$,
\begin{equation}
    t_\mathrm{synch} = \frac{3 m_e c}{4\sigma_T U_B\gamma\beta^2},
\end{equation}
to the escape time scale that we take to be of the order of the light crossing time
\begin{equation}
   t_\mathrm{esc} = \frac{r_g}{c}.
\end{equation}

If we assume that $\beta\equiv v/c\approx1$, which can be verified to be a good approximation a posteriori, we find that the cooling becomes significant around 
\begin{equation}
    \gamma_\mathrm{break} \approx 3.9\times10^3 \left(\frac{|b|}{100\:G}\right)^{-2}
\end{equation}
which gives a typical break frequency of 
\begin{equation}
    \nu_\mathrm{break} \approx 2.5\times10^{15} \left(\frac{|b|}{100\:G}\right)^{-1}\:\mathrm{Hz},
\end{equation}
so that the cooling break should be located somewhere in the UV. For $\gamma>\gamma_\mathrm{break}$  we then steepen the power-law index of the particle distribution function from $p$ to $p+1$  and the spectral index of the synchrotron emissivity in each cell by 1/2, from $(p-1)/2$ to $p/2$. In practice, we divide Eq. (\ref{eq:kappa_emiss}) by $\sqrt{\nu/\nu_\mathrm{break}}$ when $\nu>\nu_\mathrm{break}$, where $\nu_\mathrm{break}$ is the synchrotron frequency for an electron with a Lorentz factor $\gamma_\mathrm{break}$. Note that $\gamma_\mathrm{break}$ is a local property of each cell. 

We also tried to approximate the escape timescale with the accretion time scale, $r\times u^t/u^r\approx 100$ $r_g/c$.  However, in this case all the particles with $\gamma\gtrsim 10$ have enough time to cool and the power-law break is in the sub-millimeter. This would drastically reduce the emission in the X-rays. Note that a short escape timescale, comparable to the dynamical timescale, is empirically supported by the short duration of the flares ($< 1$h) in Sgr A*.

\section{Results}\label{sec:results}
In this section we present the results of our analysis, starting in \S\ref{sec:current_sheet_emission} with the current sheets properties, followed by the emitted spectrum in \S\ref{sec:spectrum} and the emitted lightcurves  in \S\ref{sec:lightcurve}.  In \S\ref{sec:MAD_vs_SANE} we carry out the same analysis on a SANE simulation, and compare the results with the MAD case.

\subsection{Emission properties of the current sheets}\label{sec:current_sheet_emission}

\begin{figure*}
\includegraphics[width=0.7\textwidth]{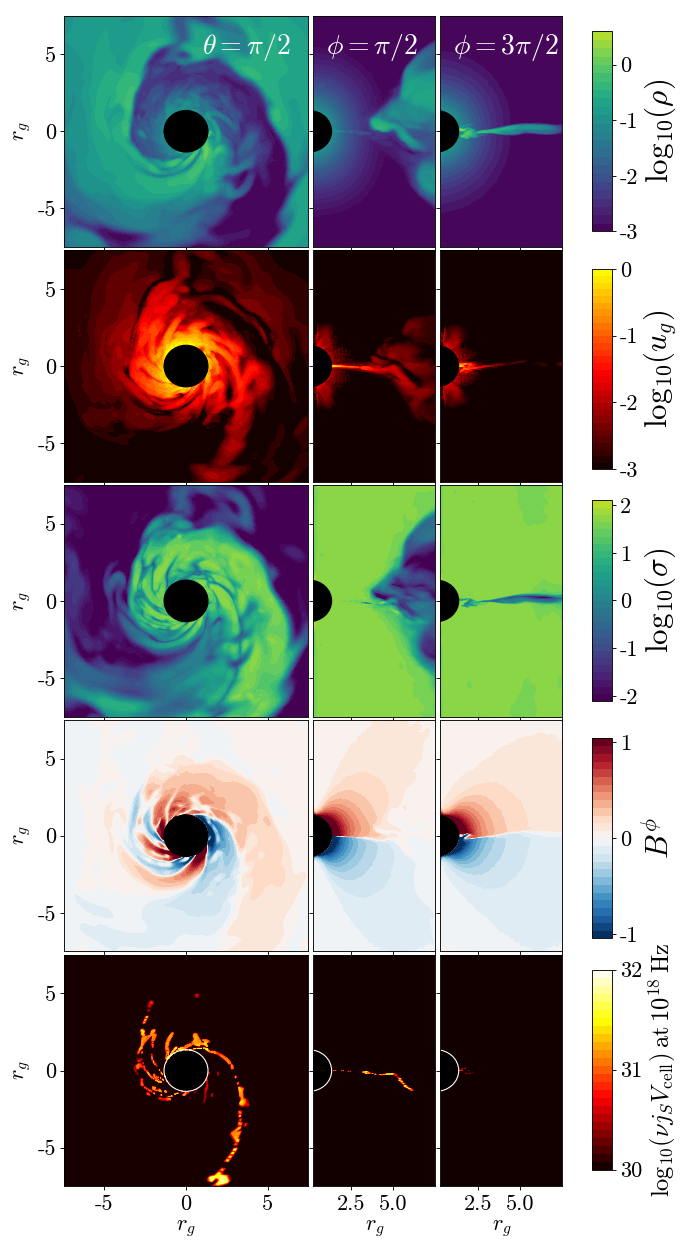}
\caption{From top to bottom: Density in code units, internal energy in code units, rest mass magnetization, 3-vector azimuthal magnetic field and emitted spectral luminosity times frequency in the X-ray at $10^{18}$ Hz. From left to right: cut in the midplane, cut at $\phi=\pi/2$ and cut at $\phi=3\pi/2$. All images are made during a flare at t=50,030 $r_g/c$.}
\label{fig:flare_rho}
\end{figure*}

\begin{figure*}
\includegraphics[width=0.7\textwidth]{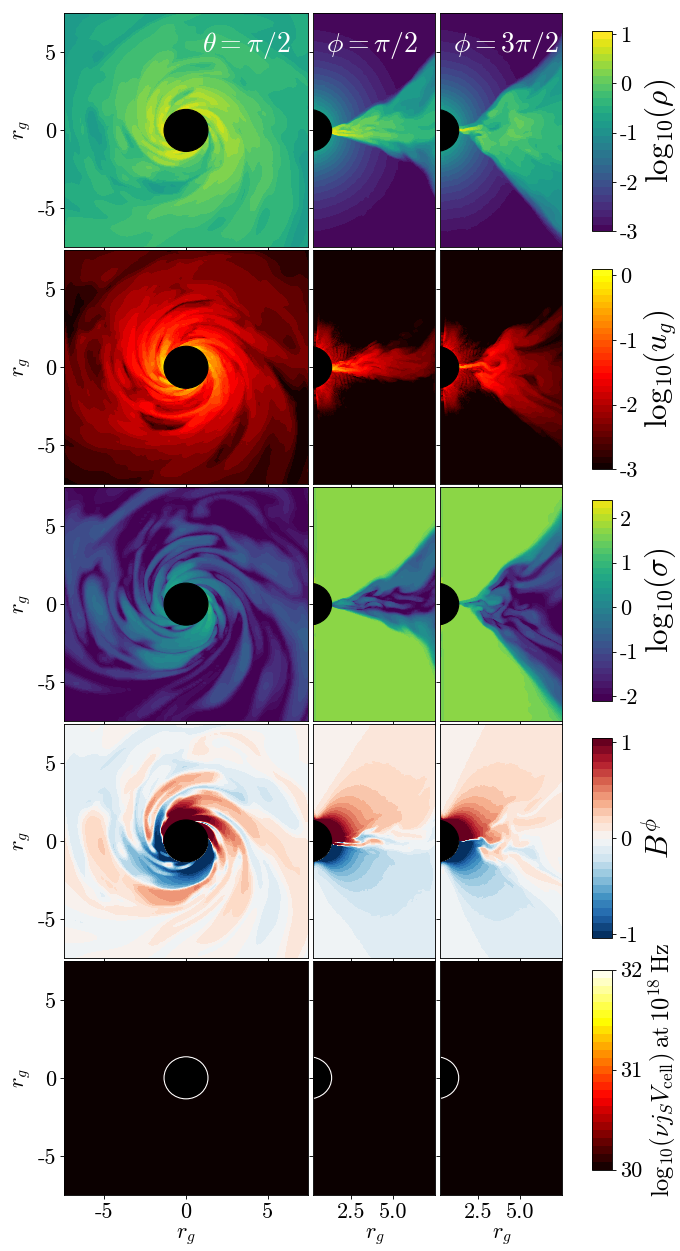}
\caption{From top to bottom: Density in code units, internal energy in code units, rest mass magnetization, 3-vector azimuthal magnetic field and emitted spectral luminosity times frequency in the X-ray at $10^{18}$ Hz. From left to right: cut in the midplane, cut at $\phi=\pi/2$ and cut at $\phi=3\pi/2$. All images are made  during a non-flaring epoch at t=36,640 $r_g/c$. }
\label{fig:noflare_rho}
\end{figure*}

\begin{figure*}
\includegraphics[width=0.95\textwidth]{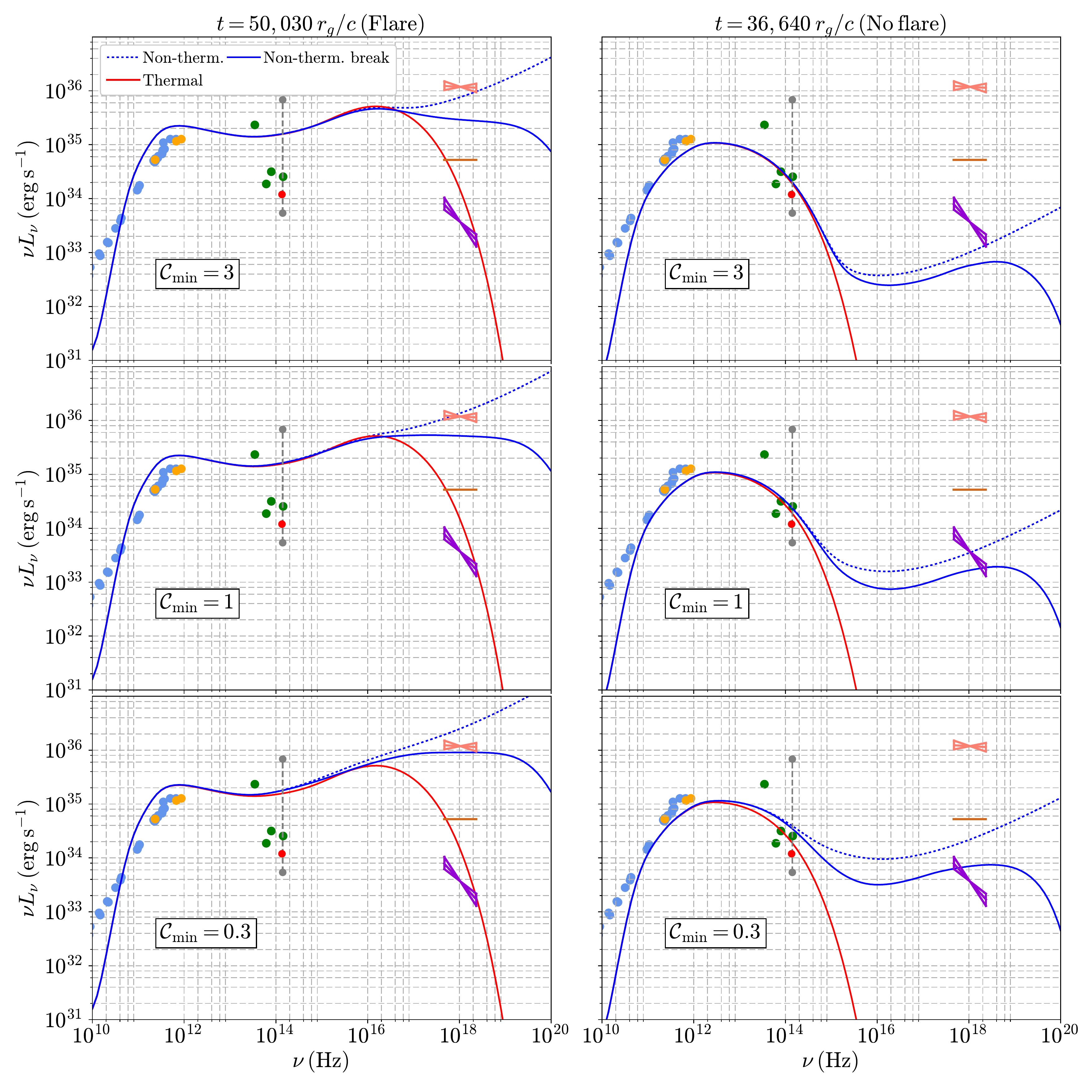}
\caption{Spectra during a flare at $t=50,030 \, r_g/c$ (left panels) and during a non-flaring epoch at t=36,640 $r_g/c$ (right panels), for different threshold $\mathcal{C}_\mathrm{min}=0.3$ (bottom), $1$ (middle) and $3$ (top). The red solid line indicates the thermal spectra, the dotted blue line the non-thermal spectra using a standard $\kappa$-distribution (without a high-energy cutoff and a cooling break) and the solid blue line the non-thermal spectra taking into account a cutoff in the particle distribution and a simple cooling break (see \S\ref{sec:cooling}). We show the radio to sub-millimeter data points from the compilation of \citet{connors2017} as light blue dots and the more recent data points from \citet{bower2019} as orange dots. We also show the mid-infrared data points from \citet{schodel2011} as green dots, the NIR maximum and minimum dereddened fluxes from \citet{do2019} as grey circles and the NIR median flux from \citet{gravity2020nir} as a red dot. Finally, we show the X-ray maximum flux ever detected by \citet{haggard2019} and quiescent flux from \citet{nowak2012} as a salmon and violet bowtie, respectively. We also show the mean flaring X-ray flux from \citet{neilsen2015} as a chocolate line.}
\label{fig:C_spectra_noflare_flare}
\end{figure*}

Figures \ref{fig:C_min_flare} and  \ref{fig:C_min_noflare} show the current sheets that are selected by our method (see \S\ref{sec:id_current_sheets}) in a MAD simulation at two different moments, corresponding to a magnetic flux eruption during which a large, magnetized bubble is expelling the flux from the black hole (Figure \ref{fig:C_min_flare}) and during a relatively quiescent period showing no eruptions (Figure \ref{fig:C_min_noflare}).  The first striking result to note is that our method selects a similar number of current sheets regardless of whether an eruption event is underway.

By comparing the bottom panels of Figures \ref{fig:flare_rho} and \ref{fig:noflare_rho}, however, which show the X-ray luminosity at $10^{18}$ Hz, we see that the emission properties of the current sheets during and outside the magnetic eruption are very different. This is mostly due to the fact that during an eruption, the current sheets are surrounded by regions of high magnetization while outside an eruption the current sheets are surrounded by regions of low magnetization. 

We can see from the poloidal cut at $\phi=3\pi/2$ on Figures \ref{fig:C_min_flare} and  \ref{fig:flare_rho} that during a magnetic eruption there is a strong current sheet forming in the region evacuated by the low-density bubble that is surrounded by material with $\sigma_\mathrm{up}\approx100$. In the same way, we can see from the equatorial cut at $\theta=\pi/2$ that the equatorial current sheet forming in the bubble is also surrounded by material with $\sigma_\mathrm{up}\approx100$. Thanks to their high upstream magnetization, these current sheets have the properties required to lead to efficient non-thermal particle acceleration and so produce significant X-ray emission. 

Contrary to the conditions during a magnetic flux eruption, a comparison of Figures \ref{fig:C_min_noflare} and \ref{fig:noflare_rho} shows that the current sheets forming in the absence of a magnetic flux eruption are surrounded by regions of much lower magnetization. This leads to very weak non-thermal particle acceleration and little X-ray emission. 

\subsection{Spectra}\label{sec:spectrum}

Figure \ref{fig:C_spectra_noflare_flare} shows two sets of spectra obtained during a magnetic eruption and during a period with no magnetic eruptions, respectively, for three different values of $\mathcal{C}_\mathrm{min}$. For both cases the red line indicates the thermal synchrotron emission, the dotted blue line the non-thermal synchrotron emission ignoring the cutoff and the solid blue line the non-thermal synchrotron emission taking into account a cutoff in the particle distribution and a simple cooling break (see \S\ref{sec:cooling}). For comparison, we show on Figure \ref{fig:C_spectra_noflare_flare} the data points from sub-millimeter to X-ray with the quiescent values as well as the maximum values ever observed at each frequency. To compare our model with the observations we normalize the flux at the sub-millimeter peak in the non-flaring case to the observed value. 

Thermal synchrotron emission is always too weak to produce the observed luminosity of X-ray flares. During a magnetic eruption, the temperature rises because of energy dissipation in the reconnecting current sheets \citep{dexter2020}, provoking a thermal flare in the infrared, optical, UV and up to the quiescent level of X-rays. Even for our most conservative value of $\mathcal{C}_\mathrm{min}=3$, the non-thermal emission extends up to the X-rays at all times (left and right panels of figure \ref{fig:C_spectra_noflare_flare}). However, the non-thermal X-ray emission is always the strongest during a magnetic flux eruption since the properties of the reconnecting current sheets are more favorable to efficient particle acceleration (see \S\ref{sec:current_sheet_emission}). 

\begin{figure}
\includegraphics[width=85mm]{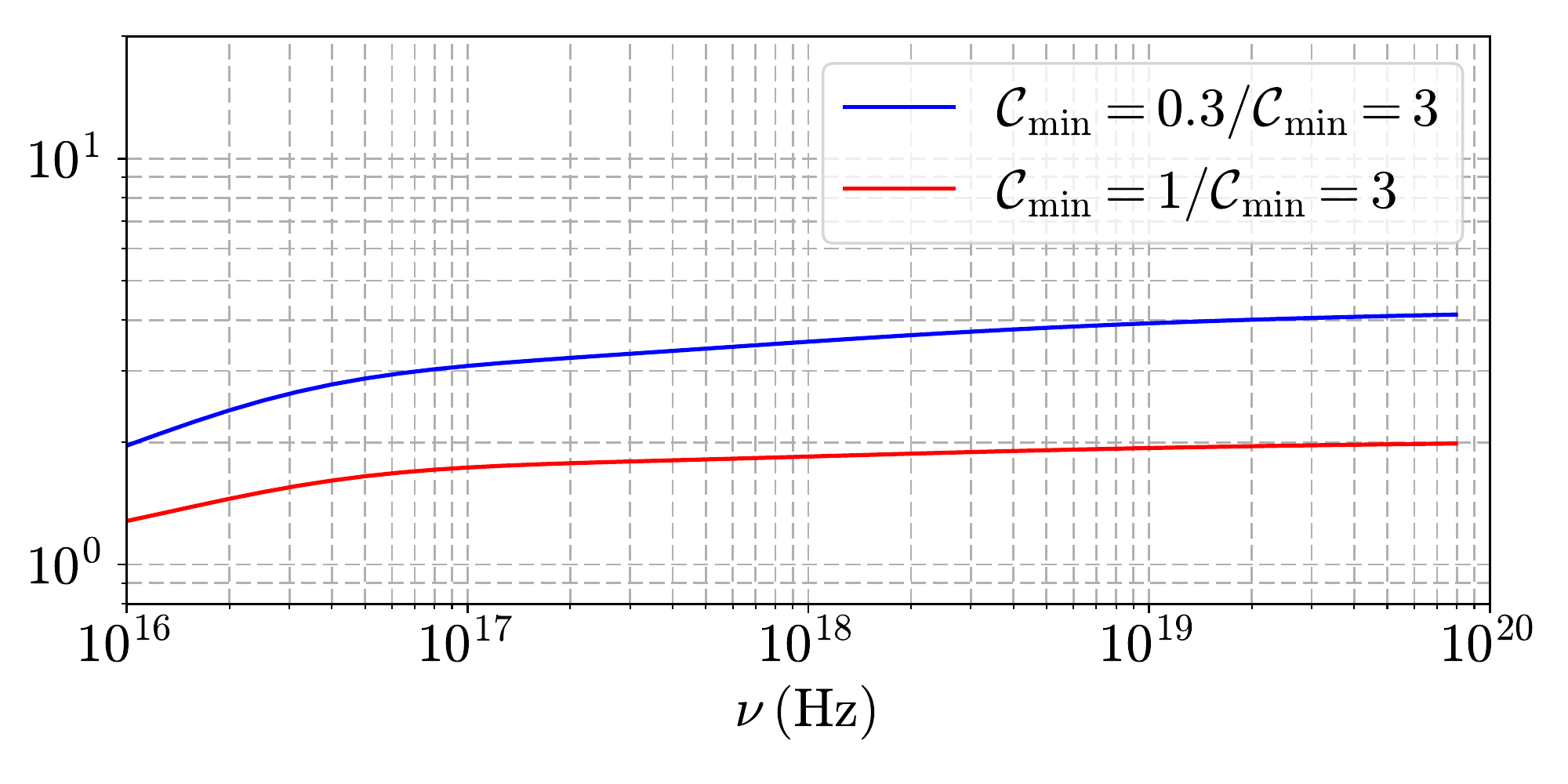}
\caption{Ratio of the high energy non-thermal spectra found using $\mathcal{C}_\mathrm{min}=0.3$ and $\mathcal{C}_\mathrm{min}=3$ (blue) and $\mathcal{C}_\mathrm{min}=1$ and $\mathcal{C}_\mathrm{min}=3$ (red) during a flare at t=50,030 $r_g/c$. This behaviour is also true outside of a flare although the ratio is larger than during a flare.}
\label{fig:comparison_C}
\end{figure}

The luminosity at high energy increases strongly with decreasing $\mathcal{C}_\mathrm{min}$ (Figure \ref{fig:C_spectra_noflare_flare}). However, we can also see from the top panel of Figure \ref{fig:comparison_C} that the shape of the spectrum at high energy does not really depend on our choice of $\mathcal{C}_\mathrm{min}$. Indeed, the top panel of Figure \ref{fig:comparison_C} shows that the ratio of the non-thermal emission in the X-rays for different values of $\mathcal{C}_\mathrm{min}$ is almost a constant at high energies. This means that when we choose lower values of  $\mathcal{C}_\mathrm{min}$ we mostly increase the surface area of the reconnecting current sheets. This can be seen on Figures \ref{fig:C_min_flare} and  \ref{fig:C_min_noflare}, where the effect of changing $\mathcal{C}_\mathrm{min}$ is primarly to increase the length and width of the reconnecting current sheets. Note that the relative amplitude of flare to quiescent luminosity increases at higher $\mathcal{C}_\mathrm{min}$.

\begin{figure*}
\includegraphics[width=\textwidth]{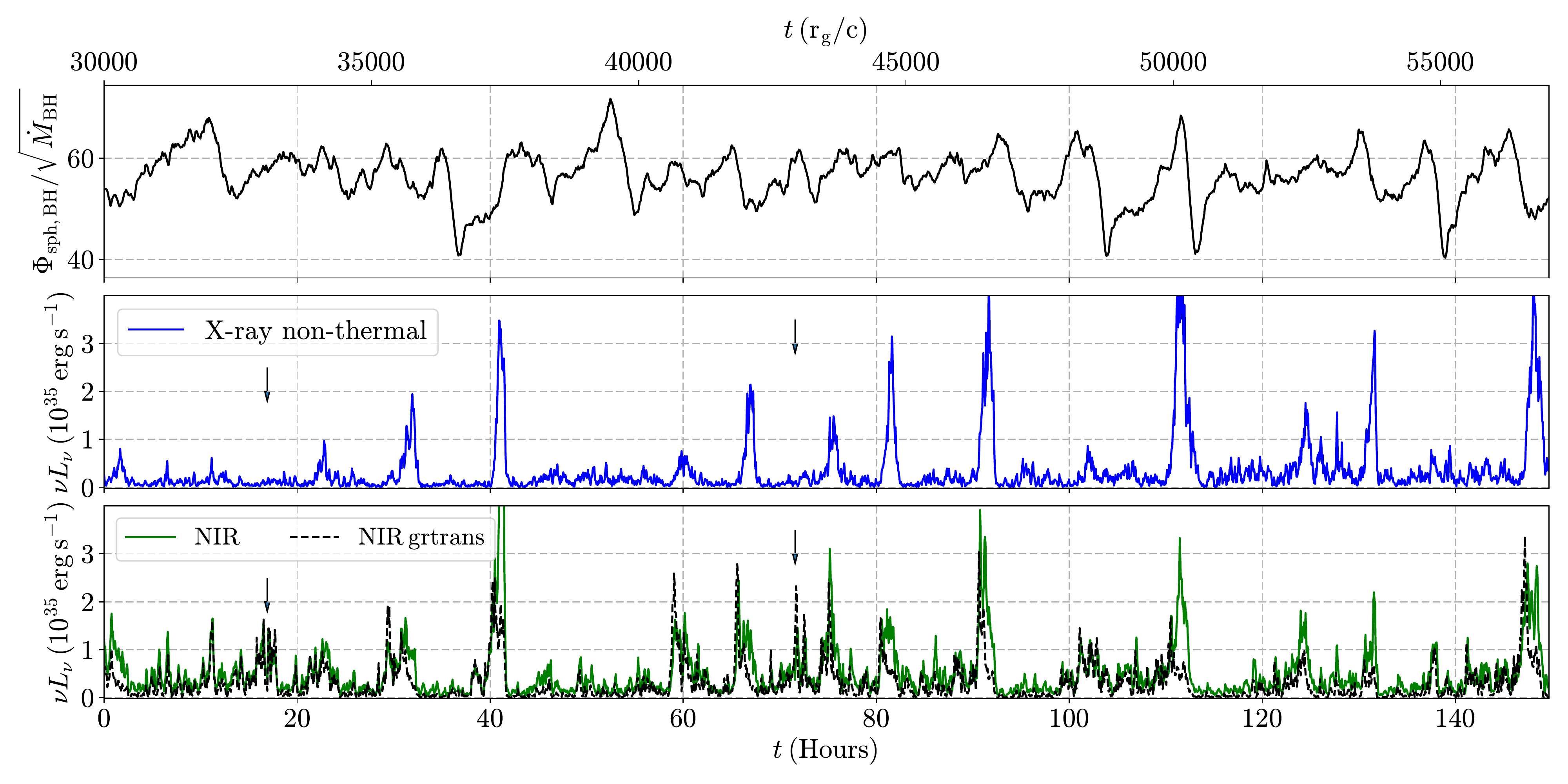}
\caption{Top panel: Normalized magnetic flux on the black hole as a function of time. Middle panel: X-ray light curve at $10^{18}$ Hz with $\mathcal{C}_\mathrm{min}=1$. Bottom panel: NIR light curve at $10^{14}$ Hz with the contribution of non-thermal electrons (solid green line) with $\mathcal{C}_\mathrm{min}=1$, compared to a fully relativistic ray tracing calculation using only thermal electrons (dashed black line). The two downward arrows at $t\approx18$ and 72 hours indicate NIR flares that do not have an X-ray flaring counterpart.}
\label{fig:LC_long}
\end{figure*} 

For $\mathcal{C}_\mathrm{min}=1$ and 3 our model satisfactorily explains the spectra of Sgr A* during and outside of a flare. Indeed, for these two values of $\mathcal{C}_\mathrm{min}$ the NIR and X-ray emission almost reach the maximum values ever observed by \cite{do2019} and \cite{haggard2019}, respectively, but also satisfy the quiescent levels of NIR and X-rays from \cite{nowak2012} and \cite{do2019}. The relatively flat X-ray spectrum during the flare is also in good agreement with the observations from \cite{haggard2019} and \cite{neilsen2015}. For $\mathcal{C}_\mathrm{min}=0.3$ the amount of X-rays in quiescence is too high to be consistent with observations.  

\subsection{Lightcurves}\label{sec:lightcurve}

Figure \ref{fig:LC_long} shows a light curve in the NIR (bottom panel) and X-ray bands (middle panel) using $\mathcal{C}_\mathrm{min}=1$,  as well as the normalized magnetic flux on the black hole \citep{tchekhovskoy2011}. For the X-ray lightcurve, we choose to use $\mathcal{C}_\mathrm{min}=1$ since this value maximizes the amount of X-rays during a flare while providing a quiescent X-ray luminosity that remains marginally consistent (though slightly too high on average) with the observations (see Figure \ref{fig:C_spectra_noflare_flare}). We note that our criterion of $\mathcal{C}_\mathrm{min}=1$ is also consistent with NIR flares and X-ray flares having roughly the same amplitude on average, in agreement with observations. We find that the flares in X-rays and NIR have a recurrence time of $\approx 10-20$ hours with amplitudes of $\approx 10$ coincident with the flux eruptions, as found in \cite{dexter2020b}. Most of the time the X-ray and NIR flares are simultaneous, especially the biggest ones, however we do see some smaller NIR flares with a weak counterpart in X-rays, as can be seen on Figure \ref{fig:LC_long} at $t\approx 18$ or 72 hours where indicated by the downward arrows. These NIR flares with a weak X-ray counterpart are due to current sheets that are not surrounded by a very magnetized medium. As we saw in \S\ref{sec:current_sheet_emission}, current sheets are forming all the time and converting magnetic energy into thermal energy to produce NIR flares but only those during magnetic flux eruptions have the required properties to create large simultaneous NIR and X-ray flares.

\begin{figure}
\includegraphics[width=80mm]{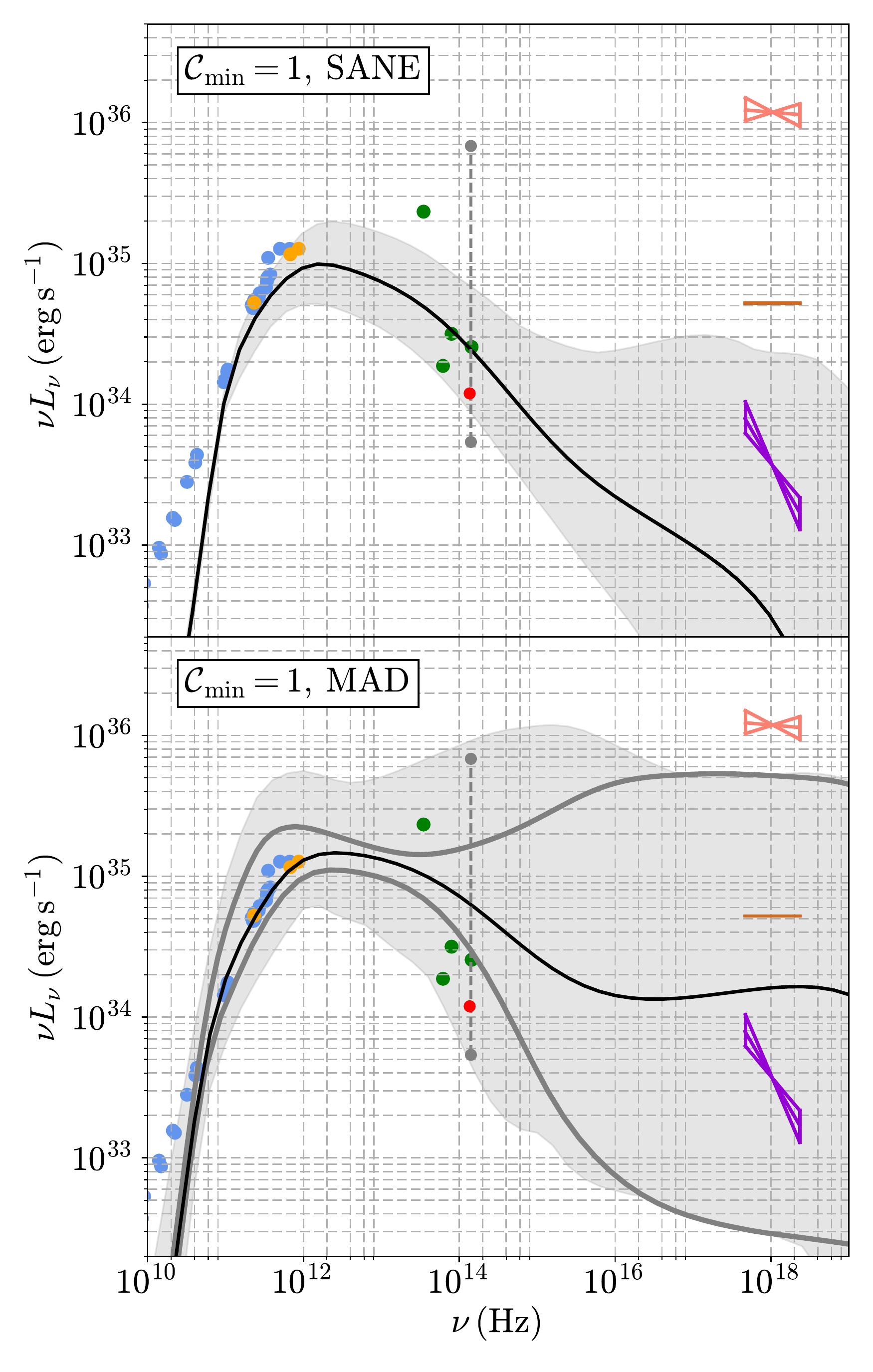}
\caption{Comparison of spectra obtained from a SANE simulation (top) and a MAD simulation (bottom) for $\mathcal{C}_\mathrm{min} =1$. The black solid line indicates the mean spectrum and the grey solid lines indicates the spectrum having the minimum and maximum X-ray luminosity at $10^{18}$ Hz. The grey shaded area indicates the range over which the entire spectrum varies and is computed by taking the minimum and maximum values at each frequency over time}. 
\label{fig:MAD_SANE_spectra}
\end{figure}

\subsection{Comparison between MAD and SANE simulation}\label{sec:MAD_vs_SANE}

Figure \ref{fig:MAD_SANE_spectra} shows a comparison of a MAD simulation (bottom panel) with a SANE simulation (top panel) with a value of $\mathcal{C}_\mathrm{min}=1$ for both. The grey shaded regions shows the minimum and maximum values in time at all frequencies. The variability in the sub-millimeter and radio is slightly larger in the MAD than the SANE model. However, we see that compared to our MAD model, the SANE model emits very little at high frequencies above the UV. This difference is due to the absence of magnetic flux eruptions in the SANE model, which are necessary to produce strong, large-scale, highly magnetized reconnecting current sheets where a large amount of energy is dumped into thermal and non-thermal particles to create the NIR and X-ray flares. Even outside of the MAD flares, we find that the SANE model has lower emission at frequencies above $\approx10^{15}$ Hz where the non-thermal emission dominates than our MAD model. This might be due to the overall higher magnetization of MAD discs compared to SANE discs, which favors non-thermal particle acceleration. Finally, we find that the mean spectra and variability obtained from our SANE simulation are in good agreement with the results of \cite{chatterjee2020} and so confirm that SANE models are not good candidates to explain the X-ray and NIR flaring behaviour of Sgr A*.

\section{Discussion and Conclusion}\label{sec:discussion}
We have used GRMHD simulations of radiatively-inefficient black hole accretion flows to identify current sheets in a realistic geometry, retrieve their physical properties, estimate their potential for producing non-thermal high-energy particles and compute the expected synchrotron emission in order to explain the X-ray flaring events in Sgr A*. We have shown that in MADs, during magnetic flux eruptions, current sheets can form in highly magnetized regions heating the plasma but also producing non-thermal high energy particles with $\gamma \lesssim 10^6$. When post-processing the synchrotron emission of the disc, we find that the deposition of thermal energy during a magnetic eruption can power NIR flares while the deposition of energy in the non-thermal particles can power X-ray flares. The NIR and X-ray emission are inherently flaring since they are associated with magnetic flux eruption events that reccur on a timescale of hours. Using a simplified synchrotron cooling prescription and a judicious choice of current sheet parameters (mainly a current-density threshold parameter $\mathcal{C}_{\rm min}$ ) we are able to reproduce, with a single set of parameters, the spectrum of Sgr A* in quiescence and in eruption from sub-millimeter to X-rays, with powerful X-ray flares reaching $5\times10^{35}$ erg/s. This result is unique to MAD simulations and is not shared with SANE simulations, which show weak non-thermal particle acceleration and so produce little X-ray variability \citep{chatterjee2020}.

One of the main caveats of our work is that we use ideal GRMHD simulations to identify reconnecting current sheets. The size and the dynamics of the current sheets is artificially governed by the finite resolution of our simulations. Hence, we cannot be certain in our quantification of the number of current sheets, their size or even the fact that they reconnect. Although the fundamental limit is the insufficient dynamic range of current global GRMHD simulations, a better quantification would be possible at higher grid resolution where reconnecting current sheets become unstable to tearing \citep{ripperda2021}.

In this work, we have used the free parameter $\mathcal{C}_\mathrm{min}$ to address the systematic uncertainty in the energetics of reconnecting current sheets in ideal GRMHD. High values of $\mathcal{C}_\mathrm{min}$ mimic a relatively low resistivity with thinner, longer and less numerous current sheets while low values of $\mathcal{C}_\mathrm{min}$ mimic a higher resistivity with larger, longer and more numerous current sheets. This method is not self-consistent, since our choice of $\mathcal{C}_\mathrm{min}$ does not influence the dynamics of the plasma. We consider it a step towards incorporating the results of kinetic reconnection studies into GRMHD simulations \citep[see also, e.g.,][]{ball2018,chatterjee2020}. We have shown that $\mathcal{C}_\mathrm{min}$ mainly acts as a normalization factor for the X-ray intensity during the flares. We find that good agreement with the observed Sgr A* X-ray luminosity in both flares and quiescence, as well as the spectral index, is possible for a value of $\mathcal{C}_\mathrm{min} \approx 1$, which is consistent with a current sheet of thickness the size of a cell at our resolution. Even a relatively modest rate of non-thermal electron acceleration in current sheets can produce the X-ray emission needed to explain the flares of Sgr A*. The very low level of the quiescent X-rays provides a strong constraint on the steady rate of energy injection.

We have also neglected relativistic effects in calculating the resulting spectra. \autoref{fig:LC_long} compares NIR light curves calculated by our method (incorporating non-thermal as well as thermal electrons) to those calculated using fully relativistic radiative transfer but only including thermal electrons \citep{dexter2020b}. Similar flaring behaviour and median luminosity are seen in both cases, with order-unity errors, although the flaring NIR emission comes from very close to the black hole. Another approximation was to neglect Compton scattering. However, we find that in our model $U_\mathrm{B}/U_\mathrm{ph}>100$ where $U_\mathrm{ph} \simeq L_\mathrm{bol}/(4\pi r_\mathrm{out}^2 c)$, with $L_\mathrm{bol}$ the bolometric luminosity from our model with $\mathcal{C}_\mathrm{min}=5$ and $r_\mathrm{out}=30\:r_g$, and $U_\mathrm{B} \simeq B_\mathrm{mean}^2$ with $B_\mathrm{mean}^2$ being the magnetic energy density averaged over a sphere of radius $r_\mathrm{out}$. This suggests that synchrotron emission is dominant over Compton scattering and justifies our approximation. Finally, to estimate the impacts of cooling on the accelerated electrons, we compared the synchrotron cooling timescale with the light crossing time scale. Obtaining a more accurate form for the distribution function would require following the histories of high-energy particles injected locally in a current sheet. Particles could leave the system very rapidly before they can efficiently cool down, or get trapped in the turbulent disc and get accelerated by diffusive processes while cooling at the same time. By following the evolution of test (tracer) particles in GRMHD simulations, such calculations should become possible in the near future.

\section*{Acknowledgements}
The authors thank the referee for their report that led to a significant improvement of our results and Dmitri Uzdensky for numerous helpful discussions and corrections on the paper. We acknowledge financial support from  NASA Astrophysics Theory Program grants NNX16AI40G, NNX17AK55G, 80NSSC20K0527, NSF Grant AST-1903335 (MCB) and an Alfred P. Sloan Research Fellowship (JD). The calculations presented here were carried out using resources supported by the NASA High-End Computing (HEC) Program through the NASA Advanced Supercomputing (NAS) Division at Ames Research Center. 

\section*{Data Availability}
The simulation data analyzed in this article will be shared on reasonable request to the corresponding author.




\bibliographystyle{mnras}
\bibliography{biblio} 

\begin{thebibliography}{}
\makeatletter
\relax
\def\mn@urlcharsother{\let\do\@makeother \do\$\do\&\do\#\do\^\do\_\do\%\do\~}
\def\mn@doi{\begingroup\mn@urlcharsother \@ifnextchar [ {\mn@doi@}
  {\mn@doi@[]}}
\def\mn@doi@[#1]#2{\def\@tempa{#1}\ifx\@tempa\@empty \href
  {http://dx.doi.org/#2} {doi:#2}\else \href {http://dx.doi.org/#2} {#1}\fi
  \endgroup}
\def\mn@eprint#1#2{\mn@eprint@#1:#2::\@nil}
\def\mn@eprint@arXiv#1{\href {http://arxiv.org/abs/#1} {{\tt arXiv:#1}}}
\def\mn@eprint@dblp#1{\href {http://dblp.uni-trier.de/rec/bibtex/#1.xml}
  {dblp:#1}}
\def\mn@eprint@#1:#2:#3:#4\@nil{\def\@tempa {#1}\def\@tempb {#2}\def\@tempc
  {#3}\ifx \@tempc \@empty \let \@tempc \@tempb \let \@tempb \@tempa \fi \ifx
  \@tempb \@empty \def\@tempb {arXiv}\fi \@ifundefined
  {mn@eprint@\@tempb}{\@tempb:\@tempc}{\expandafter \expandafter \csname
  mn@eprint@\@tempb\endcsname \expandafter{\@tempc}}}

\bibitem[\protect\citeauthoryear{{Baganoff} et~al.,}{{Baganoff}
  et~al.}{2001}]{baganoff2001}
{Baganoff} F.~K.,  et~al., 2001, \mn@doi [\nat] {10.1038/35092510}, \href
  {https://ui.adsabs.harvard.edu/abs/2001Natur.413...45B} {413, 45}

\bibitem[\protect\citeauthoryear{{Baganoff} et~al.,}{{Baganoff}
  et~al.}{2003}]{baganoff2003}
{Baganoff} F.~K.,  et~al., 2003, \mn@doi [\apj] {10.1086/375145}, \href
  {https://ui.adsabs.harvard.edu/abs/2003ApJ...591..891B} {591, 891}

\bibitem[\protect\citeauthoryear{{Balbus} \& {Hawley}}{{Balbus} \&
  {Hawley}}{1991}]{mri}
{Balbus} S.~A.,  {Hawley} J.~F.,  1991, \mn@doi [\apj] {10.1086/170270}, \href
  {http://adsabs.harvard.edu/abs/1991ApJ...376..214B} {376, 214}

\bibitem[\protect\citeauthoryear{Balbus \& Hawley}{Balbus \&
  Hawley}{1998}]{balbus98}
Balbus S.~A.,  Hawley J.~F.,  1998, Reviews of modern physics, 70, 1

\bibitem[\protect\citeauthoryear{{Balick} \& {Brown}}{{Balick} \&
  {Brown}}{1974}]{balick1974}
{Balick} B.,  {Brown} R.~L.,  1974, \mn@doi [\apj] {10.1086/153242}, \href
  {https://ui.adsabs.harvard.edu/abs/1974ApJ...194..265B} {194, 265}

\bibitem[\protect\citeauthoryear{{Ball}, {{\"O}zel}, {Psaltis}  \&
  {Chan}}{{Ball} et~al.}{2016}]{ball2016}
{Ball} D.,  {{\"O}zel} F.,  {Psaltis} D.,   {Chan} C.-k.,  2016, \mn@doi [\apj]
  {10.3847/0004-637X/826/1/77}, \href
  {https://ui.adsabs.harvard.edu/abs/2016ApJ...826...77B} {826, 77}

\bibitem[\protect\citeauthoryear{{Ball}, {Sironi}  \& {{\"O}zel}}{{Ball}
  et~al.}{2018}]{ball2018}
{Ball} D.,  {Sironi} L.,   {{\"O}zel} F.,  2018, \mn@doi [\apj]
  {10.3847/1538-4357/aac820}, \href
  {https://ui.adsabs.harvard.edu/abs/2018ApJ...862...80B} {862, 80}

\bibitem[\protect\citeauthoryear{{Bessho} \& {Bhattacharjee}}{{Bessho} \&
  {Bhattacharjee}}{2012}]{bessho2012}
{Bessho} N.,  {Bhattacharjee} A.,  2012, \mn@doi [\apj]
  {10.1088/0004-637X/750/2/129}, \href
  {https://ui.adsabs.harvard.edu/abs/2012ApJ...750..129B} {750, 129}

\bibitem[\protect\citeauthoryear{Bisnovatyi-Kogan \&
  Ruzmaikin}{Bisnovatyi-Kogan \& Ruzmaikin}{1974}]{bisnovatyi1974}
Bisnovatyi-Kogan G.,  Ruzmaikin A.,  1974, Astrophysics and Space Science, 28,
  45

\bibitem[\protect\citeauthoryear{{Blumenthal} \& {Gould}}{{Blumenthal} \&
  {Gould}}{1970}]{blumenthal1970}
{Blumenthal} G.~R.,  {Gould} R.~J.,  1970, \mn@doi [Reviews of Modern Physics]
  {10.1103/RevModPhys.42.237}, \href
  {https://ui.adsabs.harvard.edu/abs/1970RvMP...42..237B} {42, 237}

\bibitem[\protect\citeauthoryear{{Bodo}, {Tavecchio}  \& {Sironi}}{{Bodo}
  et~al.}{2021}]{bodo2021}
{Bodo} G.,  {Tavecchio} F.,   {Sironi} L.,  2021, \mn@doi [\mnras]
  {10.1093/mnras/staa3620}, \href
  {https://ui.adsabs.harvard.edu/abs/2021MNRAS.501.2836B} {501, 2836}

\bibitem[\protect\citeauthoryear{{Bower} et~al.,}{{Bower}
  et~al.}{2019}]{bower2019}
{Bower} G.~C.,  et~al., 2019, \mn@doi [\apjl] {10.3847/2041-8213/ab3397}, \href
  {https://ui.adsabs.harvard.edu/abs/2019ApJ...881L...2B} {881, L2}

\bibitem[\protect\citeauthoryear{{Cerutti}, {Werner}, {Uzdensky}  \&
  {Begelman}}{{Cerutti} et~al.}{2012}]{cerutti2012}
{Cerutti} B.,  {Werner} G.~R.,  {Uzdensky} D.~A.,   {Begelman} M.~C.,  2012,
  \mn@doi [\apjl] {10.1088/2041-8205/754/2/L33}, \href
  {https://ui.adsabs.harvard.edu/abs/2012ApJ...754L..33C} {754, L33}

\bibitem[\protect\citeauthoryear{{Cerutti}, {Werner}, {Uzdensky}  \&
  {Begelman}}{{Cerutti} et~al.}{2013}]{cerutti2013}
{Cerutti} B.,  {Werner} G.~R.,  {Uzdensky} D.~A.,   {Begelman} M.~C.,  2013,
  \mn@doi [\apj] {10.1088/0004-637X/770/2/147}, \href
  {https://ui.adsabs.harvard.edu/abs/2013ApJ...770..147C} {770, 147}

\bibitem[\protect\citeauthoryear{{Chael}, {Rowan}, {Narayan}, {Johnson}  \&
  {Sironi}}{{Chael} et~al.}{2018}]{chael2018}
{Chael} A.,  {Rowan} M.,  {Narayan} R.,  {Johnson} M.,   {Sironi} L.,  2018,
  \mn@doi [\mnras] {10.1093/mnras/sty1261}, \href
  {https://ui.adsabs.harvard.edu/abs/2018MNRAS.478.5209C} {478, 5209}

\bibitem[\protect\citeauthoryear{{Chan}, {Psaltis}, {{\"O}zel}, {Narayan}  \&
  {Sadowski}}{{Chan} et~al.}{2015a}]{chan2015}
{Chan} C.-K.,  {Psaltis} D.,  {{\"O}zel} F.,  {Narayan} R.,   {Sadowski} A.,
  2015a, \mn@doi [\apj] {10.1088/0004-637X/799/1/1}, \href
  {https://ui.adsabs.harvard.edu/\#abs/2015ApJ...799....1C} {799, 1}

\bibitem[\protect\citeauthoryear{{Chan}, {Psaltis}, {{\"O}zel}, {Medeiros},
  {Marrone}, {Sadowski}  \& {Narayan}}{{Chan} et~al.}{2015b}]{chan2015flare}
{Chan} C.-k.,  {Psaltis} D.,  {{\"O}zel} F.,  {Medeiros} L.,  {Marrone} D.,
  {Sadowski} A.,   {Narayan} R.,  2015b, \mn@doi [\apj]
  {10.1088/0004-637X/812/2/103}, \href
  {https://ui.adsabs.harvard.edu/\#abs/2015ApJ...812..103C} {812, 103}

\bibitem[\protect\citeauthoryear{Chatterjee et~al.,}{Chatterjee
  et~al.}{2020}]{chatterjee2020}
Chatterjee K.,  et~al., 2020, General relativistic MHD simulations of
  non-thermal flaring in Sagittarius A* (\mn@eprint {arXiv} {2011.08904})

\bibitem[\protect\citeauthoryear{{Connors} et~al.,}{{Connors}
  et~al.}{2017}]{connors2017}
{Connors} R.~M.~T.,  et~al., 2017, \mn@doi [\mnras] {10.1093/mnras/stw3150},
  \href {https://ui.adsabs.harvard.edu/abs/2017MNRAS.466.4121C} {466, 4121}

\bibitem[\protect\citeauthoryear{{Dexter} \& {Fragile}}{{Dexter} \&
  {Fragile}}{2013}]{dexter2013}
{Dexter} J.,  {Fragile} P.~C.,  2013, \mn@doi [\mnras] {10.1093/mnras/stt583},
  \href {https://ui.adsabs.harvard.edu/\#abs/2013MNRAS.432.2252D} {432, 2252}

\bibitem[\protect\citeauthoryear{{Dexter}, {Agol}  \& {Fragile}}{{Dexter}
  et~al.}{2009}]{dexter2009}
{Dexter} J.,  {Agol} E.,   {Fragile} P.~C.,  2009, \mn@doi [\apj]
  {10.1088/0004-637X/703/2/L142}, \href
  {https://ui.adsabs.harvard.edu/\#abs/2009ApJ...703L.142D} {703, L142}

\bibitem[\protect\citeauthoryear{{Dexter}, {Agol}, {Fragile}  \&
  {McKinney}}{{Dexter} et~al.}{2010}]{dexter2010}
{Dexter} J.,  {Agol} E.,  {Fragile} P.~C.,   {McKinney} J.~C.,  2010, \mn@doi
  [\apj] {10.1088/0004-637X/717/2/1092}, \href
  {https://ui.adsabs.harvard.edu/\#abs/2010ApJ...717.1092D} {717, 1092}

\bibitem[\protect\citeauthoryear{{Dexter} et~al.,}{{Dexter}
  et~al.}{2020a}]{dexter2020}
{Dexter} J.,  et~al., 2020a, \mn@doi [\mnras] {10.1093/mnras/staa922}, \href
  {https://ui.adsabs.harvard.edu/abs/2020MNRAS.494.4168D} {494, 4168}

\bibitem[\protect\citeauthoryear{{Dexter} et~al.,}{{Dexter}
  et~al.}{2020b}]{dexter2020b}
{Dexter} J.,  et~al., 2020b, \mn@doi [\mnras] {10.1093/mnras/staa2288}, \href
  {https://ui.adsabs.harvard.edu/abs/2020MNRAS.497.4999D} {497, 4999}

\bibitem[\protect\citeauthoryear{{Dibi}, {Drappeau}, {Fragile}, {Markoff}  \&
  {Dexter}}{{Dibi} et~al.}{2012}]{dibi2012}
{Dibi} S.,  {Drappeau} S.,  {Fragile} P.~C.,  {Markoff} S.,   {Dexter} J.,
  2012, \mn@doi [\mnras] {10.1111/j.1365-2966.2012.21857.x}, \href
  {https://ui.adsabs.harvard.edu/abs/2012MNRAS.426.1928D} {426, 1928}

\bibitem[\protect\citeauthoryear{{Dibi}, {Markoff}, {Belmont}, {Malzac},
  {Neilsen}  \& {Witzel}}{{Dibi} et~al.}{2016}]{dibi2016}
{Dibi} S.,  {Markoff} S.,  {Belmont} R.,  {Malzac} J.,  {Neilsen} J.,
  {Witzel} G.,  2016, \mn@doi [\mnras] {10.1093/mnras/stw1353}, \href
  {https://ui.adsabs.harvard.edu/abs/2016MNRAS.461..552D} {461, 552}

\bibitem[\protect\citeauthoryear{{Do} et~al.,}{{Do} et~al.}{2019}]{do2019}
{Do} T.,  et~al., 2019, \mn@doi [\apjl] {10.3847/2041-8213/ab38c3}, \href
  {https://ui.adsabs.harvard.edu/abs/2019ApJ...882L..27D} {882, L27}

\bibitem[\protect\citeauthoryear{{Dodds-Eden} et~al.,}{{Dodds-Eden}
  et~al.}{2009}]{dodds2009}
{Dodds-Eden} K.,  et~al., 2009, \mn@doi [\apj] {10.1088/0004-637X/698/1/676},
  \href {https://ui.adsabs.harvard.edu/abs/2009ApJ...698..676D} {698, 676}

\bibitem[\protect\citeauthoryear{{Doeleman} et~al.,}{{Doeleman}
  et~al.}{2008}]{doeleman2008}
{Doeleman} S.~S.,  et~al., 2008, \mn@doi [\nat] {10.1038/nature07245}, \href
  {https://ui.adsabs.harvard.edu/\#abs/2008Natur.455...78D} {455, 78}

\bibitem[\protect\citeauthoryear{{Eckart} et~al.,}{{Eckart}
  et~al.}{2006a}]{eckart2006b}
{Eckart} A.,  et~al., 2006a, \mn@doi [\aap] {10.1051/0004-6361:20054418}, \href
  {https://ui.adsabs.harvard.edu/abs/2006A&A...450..535E} {450, 535}

\bibitem[\protect\citeauthoryear{{Eckart}, {Sch{\"o}del}, {Meyer}, {Trippe},
  {Ott}  \& {Genzel}}{{Eckart} et~al.}{2006b}]{eckart2006a}
{Eckart} A.,  {Sch{\"o}del} R.,  {Meyer} L.,  {Trippe} S.,  {Ott} T.,
  {Genzel} R.,  2006b, \mn@doi [\aap] {10.1051/0004-6361:20064948}, \href
  {https://ui.adsabs.harvard.edu/abs/2006A&A...455....1E} {455, 1}

\bibitem[\protect\citeauthoryear{{Eckart} et~al.,}{{Eckart}
  et~al.}{2012}]{eckart2012}
{Eckart} A.,  et~al., 2012, \mn@doi [\aap] {10.1051/0004-6361/201117779}, \href
  {https://ui.adsabs.harvard.edu/abs/2012A&A...537A..52E} {537, A52}

\bibitem[\protect\citeauthoryear{{Fishbone} \& {Moncrief}}{{Fishbone} \&
  {Moncrief}}{1976}]{fishbone1976}
{Fishbone} L.~G.,  {Moncrief} V.,  1976, \mn@doi [\apj] {10.1086/154565}, \href
  {https://ui.adsabs.harvard.edu/abs/1976ApJ...207..962F} {207, 962}

\bibitem[\protect\citeauthoryear{{GRAVITY Collaboration} et~al.,}{{GRAVITY
  Collaboration} et~al.}{2018a}]{gravity2018a}
{GRAVITY Collaboration} et~al., 2018a, \mn@doi [A\&A]
  {10.1051/0004-6361/201833718}, 615, L15

\bibitem[\protect\citeauthoryear{{GRAVITY Collaboration} et~al.,}{{GRAVITY
  Collaboration} et~al.}{2018b}]{gravity2018b}
{GRAVITY Collaboration} et~al., 2018b, Astronomy \& Astrophysics, 618, L10

\bibitem[\protect\citeauthoryear{{Gammie}, {McKinney}  \& {T{\'o}th}}{{Gammie}
  et~al.}{2003}]{gammie2003}
{Gammie} C.~F.,  {McKinney} J.~C.,   {T{\'o}th} G.,  2003, \mn@doi [\apj]
  {10.1086/374594}, \href
  {https://ui.adsabs.harvard.edu/abs/2003ApJ...589..444G} {589, 444}

\bibitem[\protect\citeauthoryear{{Genzel}, {Sch{\"o}del}, {Ott}, {Eckart},
  {Alexander}, {Lacombe}, {Rouan}  \& {Aschenbach}}{{Genzel}
  et~al.}{2003}]{genzel2003}
{Genzel} R.,  {Sch{\"o}del} R.,  {Ott} T.,  {Eckart} A.,  {Alexander} T.,
  {Lacombe} F.,  {Rouan} D.,   {Aschenbach} B.,  2003, \mn@doi [\nat]
  {10.1038/nature02065}, \href
  {https://ui.adsabs.harvard.edu/abs/2003Natur.425..934G} {425, 934}

\bibitem[\protect\citeauthoryear{{Ghez}, {Salim}, {Hornstein}, {Tanner}, {Lu},
  {Morris}, {Becklin}  \& {Duch{\^e}ne}}{{Ghez} et~al.}{2005}]{ghez2005}
{Ghez} A.~M.,  {Salim} S.,  {Hornstein} S.~D.,  {Tanner} A.,  {Lu} J.~R.,
  {Morris} M.,  {Becklin} E.~E.,   {Duch{\^e}ne} G.,  2005, \mn@doi [\apj]
  {10.1086/427175}, \href
  {https://ui.adsabs.harvard.edu/abs/2005ApJ...620..744G} {620, 744}

\bibitem[\protect\citeauthoryear{{Gillessen} et~al.,}{{Gillessen}
  et~al.}{2017}]{gillessen2017}
{Gillessen} S.,  et~al., 2017, \mn@doi [\apj] {10.3847/1538-4357/aa5c41}, \href
  {https://ui.adsabs.harvard.edu/abs/2017ApJ...837...30G} {837, 30}

\bibitem[\protect\citeauthoryear{{Gravity Collaboration} et~al.,}{{Gravity
  Collaboration} et~al.}{2020a}]{gravity2020}
{Gravity Collaboration} et~al., 2020a, \mn@doi [\aap]
  {10.1051/0004-6361/201937233}, \href
  {https://ui.adsabs.harvard.edu/abs/2020A&A...635A.143G} {635, A143}

\bibitem[\protect\citeauthoryear{{Gravity Collaboration} et~al.,}{{Gravity
  Collaboration} et~al.}{2020b}]{gravity2020nir}
{Gravity Collaboration} et~al., 2020b, \mn@doi [\aap]
  {10.1051/0004-6361/202037717}, \href
  {https://ui.adsabs.harvard.edu/abs/2020A&A...638A...2G} {638, A2}

\bibitem[\protect\citeauthoryear{{Gravity Collaboration} et~al.,}{{Gravity
  Collaboration} et~al.}{2020c}]{gravity2020pol}
{Gravity Collaboration} et~al., 2020c, \mn@doi [\aap]
  {10.1051/0004-6361/202038283}, \href
  {https://ui.adsabs.harvard.edu/abs/2020A&A...643A..56G} {643, A56}

\bibitem[\protect\citeauthoryear{{Guo}, {Li}, {Daughton}  \& {Liu}}{{Guo}
  et~al.}{2014}]{guo2014}
{Guo} F.,  {Li} H.,  {Daughton} W.,   {Liu} Y.-H.,  2014, \mn@doi [\prl]
  {10.1103/PhysRevLett.113.155005}, \href
  {https://ui.adsabs.harvard.edu/abs/2014PhRvL.113o5005G} {113, 155005}

\bibitem[\protect\citeauthoryear{{Guo} et~al.,}{{Guo} et~al.}{2016}]{guo2016}
{Guo} F.,  et~al., 2016, \mn@doi [\apjl] {10.3847/2041-8205/818/1/L9}, \href
  {https://ui.adsabs.harvard.edu/abs/2016ApJ...818L...9G} {818, L9}

\bibitem[\protect\citeauthoryear{{Haggard} et~al.,}{{Haggard}
  et~al.}{2019}]{haggard2019}
{Haggard} D.,  et~al., 2019, \mn@doi [\apj] {10.3847/1538-4357/ab4a7f}, \href
  {https://ui.adsabs.harvard.edu/abs/2019ApJ...886...96H} {886, 96}

\bibitem[\protect\citeauthoryear{{Hakobyan}, {Petropoulou}, {Spitkovsky}  \&
  {Sironi}}{{Hakobyan} et~al.}{2021}]{hakobyan2021}
{Hakobyan} H.,  {Petropoulou} M.,  {Spitkovsky} A.,   {Sironi} L.,  2021,
  \mn@doi [\apj] {10.3847/1538-4357/abedac}, \href
  {https://ui.adsabs.harvard.edu/abs/2021ApJ...912...48H} {912, 48}

\bibitem[\protect\citeauthoryear{{Hirose}, {Krolik}, {De Villiers}  \&
  {Hawley}}{{Hirose} et~al.}{2004}]{hirose2004}
{Hirose} S.,  {Krolik} J.~H.,  {De Villiers} J.-P.,   {Hawley} J.~F.,  2004,
  \mn@doi [\apj] {10.1086/383184}, \href
  {https://ui.adsabs.harvard.edu/abs/2004ApJ...606.1083H} {606, 1083}

\bibitem[\protect\citeauthoryear{{Howes}}{{Howes}}{2010}]{howes2010}
{Howes} G.~G.,  2010, \mn@doi [\mnras] {10.1111/j.1745-3933.2010.00958.x},
  \href {https://ui.adsabs.harvard.edu/abs/2010MNRAS.409L.104H} {409, L104}

\bibitem[\protect\citeauthoryear{{Igumenshchev}}{{Igumenshchev}}{2008}]{igumenshchev2008}
{Igumenshchev} I.~V.,  2008, \mn@doi [\apj] {10.1086/529025}, \href
  {https://ui.adsabs.harvard.edu/abs/2008ApJ...677..317I} {677, 317}

\bibitem[\protect\citeauthoryear{{Kagan}, {Milosavljevi{\'c}}  \&
  {Spitkovsky}}{{Kagan} et~al.}{2013}]{kagan2013}
{Kagan} D.,  {Milosavljevi{\'c}} M.,   {Spitkovsky} A.,  2013, \mn@doi [\apj]
  {10.1088/0004-637X/774/1/41}, \href
  {https://ui.adsabs.harvard.edu/abs/2013ApJ...774...41K} {774, 41}

\bibitem[\protect\citeauthoryear{{Kawazura}, {Barnes}  \&
  {Schekochihin}}{{Kawazura} et~al.}{2019}]{kawazura2019}
{Kawazura} Y.,  {Barnes} M.,   {Schekochihin} A.~A.,  2019, \mn@doi
  [Proceedings of the National Academy of Science] {10.1073/pnas.1812491116},
  \href {https://ui.adsabs.harvard.edu/abs/2019PNAS..116..771K} {116, 771}

\bibitem[\protect\citeauthoryear{{Markoff}, {Falcke}, {Yuan}  \&
  {Biermann}}{{Markoff} et~al.}{2001}]{markoff2001}
{Markoff} S.,  {Falcke} H.,  {Yuan} F.,   {Biermann} P.~L.,  2001, \mn@doi
  [\aap] {10.1051/0004-6361:20011346}, \href
  {https://ui.adsabs.harvard.edu/abs/2001A&A...379L..13M} {379, L13}

\bibitem[\protect\citeauthoryear{{Marrone} et~al.,}{{Marrone}
  et~al.}{2008}]{marrone2008}
{Marrone} D.~P.,  et~al., 2008, \mn@doi [\apj] {10.1086/588806}, \href
  {https://ui.adsabs.harvard.edu/abs/2008ApJ...682..373M} {682, 373}

\bibitem[\protect\citeauthoryear{Mayer-Hasselwander et~al.,}{Mayer-Hasselwander
  et~al.}{1998}]{mayer1998}
Mayer-Hasselwander H.,  et~al., 1998, Astronomy and Astrophysics, 335, 161

\bibitem[\protect\citeauthoryear{{Mo{\'s}cibrodzka}, {Gammie}, {Dolence},
  {Shiokawa}  \& {Leung}}{{Mo{\'s}cibrodzka} et~al.}{2009}]{moscibrodzka2009}
{Mo{\'s}cibrodzka} M.,  {Gammie} C.~F.,  {Dolence} J.~C.,  {Shiokawa} H.,
  {Leung} P.~K.,  2009, \mn@doi [\apj] {10.1088/0004-637X/706/1/497}, \href
  {https://ui.adsabs.harvard.edu/\#abs/2009ApJ...706..497M} {706, 497}

\bibitem[\protect\citeauthoryear{{Neilsen} et~al.,}{{Neilsen}
  et~al.}{2013}]{neilsen2013}
{Neilsen} J.,  et~al., 2013, \mn@doi [\apj] {10.1088/0004-637X/774/1/42}, \href
  {https://ui.adsabs.harvard.edu/abs/2013ApJ...774...42N} {774, 42}

\bibitem[\protect\citeauthoryear{{Neilsen} et~al.,}{{Neilsen}
  et~al.}{2015}]{neilsen2015}
{Neilsen} J.,  et~al., 2015, \mn@doi [\apj] {10.1088/0004-637X/799/2/199},
  \href {https://ui.adsabs.harvard.edu/abs/2015ApJ...799..199N} {799, 199}

\bibitem[\protect\citeauthoryear{{Nowak} et~al.,}{{Nowak}
  et~al.}{2012}]{nowak2012}
{Nowak} M.~A.,  et~al., 2012, \mn@doi [\apj] {10.1088/0004-637X/759/2/95},
  \href {https://ui.adsabs.harvard.edu/abs/2012ApJ...759...95N} {759, 95}

\bibitem[\protect\citeauthoryear{{Pandya}, {Zhang}, {Chandra}  \&
  {Gammie}}{{Pandya} et~al.}{2016}]{pandya2016}
{Pandya} A.,  {Zhang} Z.,  {Chandra} M.,   {Gammie} C.~F.,  2016, \mn@doi
  [\apj] {10.3847/0004-637X/822/1/34}, \href
  {https://ui.adsabs.harvard.edu/abs/2016ApJ...822...34P} {822, 34}

\bibitem[\protect\citeauthoryear{{Petersen} \& {Gammie}}{{Petersen} \&
  {Gammie}}{2020}]{petersen2020}
{Petersen} E.,  {Gammie} C.,  2020, \mn@doi [\mnras] {10.1093/mnras/staa826},
  \href {https://ui.adsabs.harvard.edu/abs/2020MNRAS.494.5923P} {494, 5923}

\bibitem[\protect\citeauthoryear{{Petropoulou} \& {Sironi}}{{Petropoulou} \&
  {Sironi}}{2018}]{petropoulou2018}
{Petropoulou} M.,  {Sironi} L.,  2018, \mn@doi [\mnras]
  {10.1093/mnras/sty2702}, \href
  {https://ui.adsabs.harvard.edu/abs/2018MNRAS.481.5687P} {481, 5687}

\bibitem[\protect\citeauthoryear{{Ponti} et~al.,}{{Ponti}
  et~al.}{2017}]{ponti2017}
{Ponti} G.,  et~al., 2017, \mn@doi [\mnras] {10.1093/mnras/stx596}, \href
  {https://ui.adsabs.harvard.edu/abs/2017MNRAS.468.2447P} {468, 2447}

\bibitem[\protect\citeauthoryear{{Porth}, {Mizuno}, {Younsi}  \&
  {Fromm}}{{Porth} et~al.}{2021}]{porth2021}
{Porth} O.,  {Mizuno} Y.,  {Younsi} Z.,   {Fromm} C.~M.,  2021, \mn@doi
  [\mnras] {10.1093/mnras/stab163}, \href
  {https://ui.adsabs.harvard.edu/abs/2021MNRAS.502.2023P} {502, 2023}

\bibitem[\protect\citeauthoryear{{Quataert}}{{Quataert}}{2002}]{quataert2002}
{Quataert} E.,  2002, \mn@doi [\apj] {10.1086/341425}, \href
  {https://ui.adsabs.harvard.edu/abs/2002ApJ...575..855Q} {575, 855}

\bibitem[\protect\citeauthoryear{{Ressler}, {Tchekhovskoy}, {Quataert},
  {Chandra}  \& {Gammie}}{{Ressler} et~al.}{2015}]{ressler2015}
{Ressler} S.~M.,  {Tchekhovskoy} A.,  {Quataert} E.,  {Chandra} M.,   {Gammie}
  C.~F.,  2015, \mn@doi [\mnras] {10.1093/mnras/stv2084}, \href
  {https://ui.adsabs.harvard.edu/abs/2015MNRAS.454.1848R} {454, 1848}

\bibitem[\protect\citeauthoryear{{Ressler}, {Tchekhovskoy}, {Quataert}  \&
  {Gammie}}{{Ressler} et~al.}{2017}]{ressler2017}
{Ressler} S.~M.,  {Tchekhovskoy} A.,  {Quataert} E.,   {Gammie} C.~F.,  2017,
  \mn@doi [\mnras] {10.1093/mnras/stx364}, \href
  {https://ui.adsabs.harvard.edu/\#abs/2017MNRAS.467.3604R} {467, 3604}

\bibitem[\protect\citeauthoryear{{Ripperda}, {Bacchini}  \&
  {Philippov}}{{Ripperda} et~al.}{2020}]{ripperda2020}
{Ripperda} B.,  {Bacchini} F.,   {Philippov} A.~A.,  2020, \mn@doi [\apj]
  {10.3847/1538-4357/ababab}, \href
  {https://ui.adsabs.harvard.edu/abs/2020ApJ...900..100R} {900, 100}

\bibitem[\protect\citeauthoryear{{Ripperda}, {Liska}, {Chatterjee}, {Musoke},
  {Philippov}, {Markoff}, {Tchekhovskoy}  \& {Younsi}}{{Ripperda}
  et~al.}{2021}]{ripperda2021}
{Ripperda} B.,  {Liska} M.,  {Chatterjee} K.,  {Musoke} G.,  {Philippov} A.~A.,
   {Markoff} S.~B.,  {Tchekhovskoy} A.,   {Younsi} Z.,  2021, arXiv e-prints,
  \href {https://ui.adsabs.harvard.edu/abs/2021arXiv210915115R} {p.
  arXiv:2109.15115}

\bibitem[\protect\citeauthoryear{{Rowan}, {Sironi}  \& {Narayan}}{{Rowan}
  et~al.}{2017}]{rowan2017}
{Rowan} M.~E.,  {Sironi} L.,   {Narayan} R.,  2017, \mn@doi [\apj]
  {10.3847/1538-4357/aa9380}, \href
  {https://ui.adsabs.harvard.edu/abs/2017ApJ...850...29R} {850, 29}

\bibitem[\protect\citeauthoryear{{Ryan}, {Ressler}, {Dolence}, {Tchekhovskoy},
  {Gammie}  \& {Quataert}}{{Ryan} et~al.}{2017}]{ryan2017}
{Ryan} B.~R.,  {Ressler} S.~M.,  {Dolence} J.~C.,  {Tchekhovskoy} A.,  {Gammie}
  C.,   {Quataert} E.,  2017, \mn@doi [\apjl] {10.3847/2041-8213/aa8034}, \href
  {https://ui.adsabs.harvard.edu/abs/2017ApJ...844L..24R} {844, L24}

\bibitem[\protect\citeauthoryear{{Sch{\"o}del}, {Morris}, {Muzic}, {Alberdi},
  {Meyer}, {Eckart}  \& {Gezari}}{{Sch{\"o}del} et~al.}{2011}]{schodel2011}
{Sch{\"o}del} R.,  {Morris} M.~R.,  {Muzic} K.,  {Alberdi} A.,  {Meyer} L.,
  {Eckart} A.,   {Gezari} D.~Y.,  2011, \mn@doi [\aap]
  {10.1051/0004-6361/201116994}, \href
  {https://ui.adsabs.harvard.edu/abs/2011A&A...532A..83S} {532, A83}

\bibitem[\protect\citeauthoryear{{Shcherbakov}, {Penna}  \&
  {McKinney}}{{Shcherbakov} et~al.}{2012}]{shcherbakov2012}
{Shcherbakov} R.~V.,  {Penna} R.~F.,   {McKinney} J.~C.,  2012, \mn@doi [\apj]
  {10.1088/0004-637X/755/2/133}, \href
  {https://ui.adsabs.harvard.edu/\#abs/2012ApJ...755..133S} {755, 133}

\bibitem[\protect\citeauthoryear{{Sironi} \& {Spitkovsky}}{{Sironi} \&
  {Spitkovsky}}{2014}]{sironi2014}
{Sironi} L.,  {Spitkovsky} A.,  2014, \mn@doi [\apjl]
  {10.1088/2041-8205/783/1/L21}, \href
  {https://ui.adsabs.harvard.edu/abs/2014ApJ...783L..21S} {783, L21}

\bibitem[\protect\citeauthoryear{{Tchekhovskoy}}{{Tchekhovskoy}}{2019}]{tchekhovskoy2019}
{Tchekhovskoy} A.,  2019, {HARMPI: 3D massively parallel general relativictic
  MHD code} (\mn@eprint {ascl} {1912.014})

\bibitem[\protect\citeauthoryear{{Tchekhovskoy}, {Narayan}  \&
  {McKinney}}{{Tchekhovskoy} et~al.}{2011}]{tchekhovskoy2011}
{Tchekhovskoy} A.,  {Narayan} R.,   {McKinney} J.~C.,  2011, \mn@doi [\mnras]
  {10.1111/j.1745-3933.2011.01147.x}, \href
  {https://ui.adsabs.harvard.edu/abs/2011MNRAS.418L..79T} {418, L79}

\bibitem[\protect\citeauthoryear{{Uzdensky}, {Cerutti}  \&
  {Begelman}}{{Uzdensky} et~al.}{2011}]{uzdensky2011}
{Uzdensky} D.~A.,  {Cerutti} B.,   {Begelman} M.~C.,  2011, \mn@doi [\apjl]
  {10.1088/2041-8205/737/2/L40}, \href
  {https://ui.adsabs.harvard.edu/abs/2011ApJ...737L..40U} {737, L40}

\bibitem[\protect\citeauthoryear{{Viergutz}}{{Viergutz}}{1993}]{viergutz1993}
{Viergutz} S.~U.,  1993, \aap, \href
  {https://ui.adsabs.harvard.edu/abs/1993A&A...272..355V} {272, 355}

\bibitem[\protect\citeauthoryear{{Werner}, {Uzdensky}, {Cerutti}, {Nalewajko}
  \& {Begelman}}{{Werner} et~al.}{2016}]{werner2016}
{Werner} G.~R.,  {Uzdensky} D.~A.,  {Cerutti} B.,  {Nalewajko} K.,   {Begelman}
  M.~C.,  2016, \mn@doi [\apjl] {10.3847/2041-8205/816/1/L8}, \href
  {https://ui.adsabs.harvard.edu/abs/2016ApJ...816L...8W} {816, L8}

\bibitem[\protect\citeauthoryear{{Werner}, {Uzdensky}, {Begelman}, {Cerutti}
  \& {Nalewajko}}{{Werner} et~al.}{2018}]{werner2018}
{Werner} G.~R.,  {Uzdensky} D.~A.,  {Begelman} M.~C.,  {Cerutti} B.,
  {Nalewajko} K.,  2018, \mn@doi [\mnras] {10.1093/mnras/stx2530}, \href
  {https://ui.adsabs.harvard.edu/abs/2018MNRAS.473.4840W} {473, 4840}

\bibitem[\protect\citeauthoryear{{Werner}, {Philippov}  \& {Uzdensky}}{{Werner}
  et~al.}{2019}]{werner2019}
{Werner} G.~R.,  {Philippov} A.~A.,   {Uzdensky} D.~A.,  2019, \mn@doi [\mnras]
  {10.1093/mnrasl/sly157}, \href
  {https://ui.adsabs.harvard.edu/abs/2019MNRAS.482L..60W} {482, L60}

\bibitem[\protect\citeauthoryear{{White} \& {Quataert}}{{White} \&
  {Quataert}}{2021}]{white2021}
{White} C.~J.,  {Quataert} E.,  2021, arXiv e-prints, \href
  {https://ui.adsabs.harvard.edu/abs/2021arXiv210412896W} {p. arXiv:2104.12896}

\bibitem[\protect\citeauthoryear{{Yuan} \& {Narayan}}{{Yuan} \&
  {Narayan}}{2014}]{yuan2014}
{Yuan} F.,  {Narayan} R.,  2014, \mn@doi [\araa]
  {10.1146/annurev-astro-082812-141003}, \href
  {https://ui.adsabs.harvard.edu/abs/2014ARA&A..52..529Y} {52, 529}

\bibitem[\protect\citeauthoryear{{Yusef-Zadeh}, {Wardle}, {Heinke}, {Dowell},
  {Roberts}, {Baganoff}  \& {Cotton}}{{Yusef-Zadeh}
  et~al.}{2008}]{yusefzadeh2008}
{Yusef-Zadeh} F.,  {Wardle} M.,  {Heinke} C.,  {Dowell} C.~D.,  {Roberts} D.,
  {Baganoff} F.~K.,   {Cotton} W.,  2008, \mn@doi [\apj] {10.1086/588803},
  \href {https://ui.adsabs.harvard.edu/abs/2008ApJ...682..361Y} {682, 361}

\bibitem[\protect\citeauthoryear{{Zhdankin}, {Uzdensky}, {Perez}  \&
  {Boldyrev}}{{Zhdankin} et~al.}{2013}]{zhdankin2013}
{Zhdankin} V.,  {Uzdensky} D.~A.,  {Perez} J.~C.,   {Boldyrev} S.,  2013,
  \mn@doi [\apj] {10.1088/0004-637X/771/2/124}, \href
  {https://ui.adsabs.harvard.edu/abs/2013ApJ...771..124Z} {771, 124}

\makeatother
\end{thebibliography}


\bsp	
\label{lastpage}
\end{document}